\documentclass[12pt, draftclsnofoot,onecolumn]{IEEEtran}

\usepackage{graphicx,subcaption,xcolor,setspace,amsthm,cite,booktabs,epstopdf,dblfloatfix,algorithm,algcompatible,blindtext}
\usepackage[binary-units,detect-weight=true, detect-family=true]{siunitx}
\DeclareSIUnit \bitspersecond {bps}
\newcommand{\centrallocation}{/Users/AnumAli/Dropbox/STUDY/Central} 
\usepackage{amsmath,amssymb}

\newcommand{\bbE}{{\mathbb{E}}}

\newcommand{\bbR}{{\mathbb{R}}}

\newcommand{\ba}{{\mathbf{a}}}

\newcommand{\bff}{{\mathbf{f}}}
\newcommand{\bg}{{\mathbf{g}}}

\newcommand{\bp}{{\mathbf{p}}}
\newcommand{\bq}{{\mathbf{q}}}

\newcommand{\bt}{{\mathbf{t}}}

\newcommand{\bv}{{\mathbf{v}}}

\newcommand{\bx}{{\mathbf{x}}}

\newcommand{\bzero}{{\mathbf{0}}}
\newcommand{\bone}{{\mathbf{1}}}
\newcommand{\bA}{{\mathbf{A}}}

\newcommand{\bD}{{\mathbf{D}}}

\newcommand{\bF}{{\mathbf{F}}}
\newcommand{\bG}{{\mathbf{G}}}
\newcommand{\bH}{{\mathbf{H}}}
\newcommand{\bI}{{\mathbf{I}}}

\newcommand{\bN}{{\mathbf{N}}}

\newcommand{\bQ}{{\mathbf{Q}}}
\newcommand{\bR}{{\mathbf{R}}}

\newcommand{\bT}{{\mathbf{T}}}

\newcommand{\bV}{{\mathbf{V}}}

\newcommand{\bX}{{\mathbf{X}}}
\newcommand{\bY}{{\mathbf{Y}}}

\newcommand{\rmc}{{\mathrm{c}}}

\newcommand{\rmi}{{\mathrm{i}}}

\newcommand{\rml}{{\mathrm{l}}}
\newcommand{\rmm}{{\mathrm{m}}}
\newcommand{\rmn}{{\mathrm{n}}}
\newcommand{\rmo}{{\mathrm{o}}}
\newcommand{\rmp}{{\mathrm{p}}}

\newcommand{\rmr}{{\mathrm{r}}}
\newcommand{\rms}{{\mathrm{s}}}
\newcommand{\rmt}{{\mathrm{t}}}

\newcommand{\rmw}{{\mathrm{w}}}

\newcommand{\rmF}{{\mathrm{F}}}

\newcommand{\rmR}{{\mathrm{R}}}

\newcommand{\rmT}{{\mathrm{T}}}

\newcommand{\rmX}{{\mathrm{X}}}

\newcommand{\cB}{\mathcal{B}}
\newcommand{\cC}{\mathcal{C}}

\newcommand{\cJ}{\mathcal{J}}

\newcommand{\cM}{\mathcal{M}}
\newcommand{\cN}{\mathcal{N}}

\newcommand{\cR}{\mathcal{R}}

\newcommand{\cX}{\mathcal{X}}

\newcommand{\sfs}{{\mathsf{s}}}

\newcommand{\sfy}{{\mathsf{y}}}

\newcommand{\bsfg}{\boldsymbol{\mathsf{g}}}

\newcommand{\bsfv}{\boldsymbol{\mathsf{v}}}

\newcommand{\bsfy}{\boldsymbol{\mathsf{y}}}

\newcommand{\bsfG}{\boldsymbol{\mathsf{G}}}
\newcommand{\bsfH}{\boldsymbol{\mathsf{H}}}

\newcommand{\bsfV}{\boldsymbol{\mathsf{V}}}

\newcommand{\bsfY}{\boldsymbol{\mathsf{Y}}}

\newcommand{\bPsi}{\boldsymbol{\Psi}}

\newcommand{\transp}{{\sf T}}
\newcommand{\compj}{{\rm j}}

\renewcommand{\vec}{{\rm vec}}

\makeatletter
\def\munderbar#1{\underline{\sbox\tw@{$#1$}\dp\tw@\z@\box\tw@}}
\makeatother

\algnewcommand\INPUT{\item[\textbf{Input:}]}
\algnewcommand\OUTPUT{\item[\textbf{Output:}]}
\algnewcommand\OFFLINE{\item[\textbf{Offline Calculation:}]}
\newcommand{\ul}{\underline}
\newcommand{\subsGHz}{sub-6 \GHz}

\newcommand{\SubsGHz}{Sub-6 \GHz}
\newcommand{\GHz}{\SI{}{\giga\hertz}}
\newcommand{\MHz}{\SI{}{\mega\hertz}}

\newcommand{\dBm}{\SI{}{\decibel}\rmm}
\newcommand{\ns}{\SI{}{\nano\second}}
\newcommand{\SNR}{\rm{SNR}}
\newcommand{\MRX}{M_{\rmR\rmX}}
\newcommand{\MTX}{M_{\rmT\rmX}}
\newcommand{\NRX}{N_{\rmR\rmX}}
\newcommand{\NTX}{N_{\rmT\rmX}}
\newcommand{\DRX}{D_{\rmR\rmX}}
\newcommand{\DTX}{D_{\rmT\rmX}}
\newcommand{\Lc}{L_{\rmc}}
\newcommand{\rhopl}{\rho_{\rmp\rml}}
\newcommand{\aRX}{\ba_{\rmR\rmX}}
\newcommand{\aTX}{\ba_{\rmT\rmX}}
\newcommand{\ARX}{\bA_{\rmR\rmX}}
\newcommand{\ATX}{\bA_{\rmT\rmX}}

\newcommand{\noisepsf}{\sigma_{\check\bsfv}^2}

\newcommand{\MRXg}{\ul{M}_{\rmR\rmX}}
\newcommand{\MTXg}{\ul{M}_{\rmT\rmX}}

\newcommand{\Ntrg}{\ul{N}_{\rmt\rmr}}

\newcommand{\aRXg}{\ul{\ba}_{\rmR\rmX}}
\newcommand{\aTXg}{\ul{\ba}_{\rmT\rmX}}
\newcommand{\ARXg}{\ul{\bA}_{\rmR\rmX}}
\newcommand{\ATXg}{\ul{\bA}_{\rmT\rmX}}
\newcommand{\noisepg}{\sigma_{\ul{\bv}}^2}

\newcommand{\NTXb}{\bar{N}_{\rmT\rmX}}
\newcommand{\bHg}{\ul{\bH}}

\newcommand{\bRg}{\ul{\bR}}
\newcommand{\bGg}{\ul{\bG}}
\newcommand{\bgg}{\ul{\bg}}

\newcommand{\Cg}{\ul{C}}

\newcommand{\cg}{\ul{c}}
\newcommand{\btg}{\ul{\bt}}

\newcommand{\fg}{\ul{f}}
\newcommand{\Pt}{P_{\rmt}}
\newcommand{\Ptg}{\ul{P} _{\rmt}}

\newcommand{\Ts}{T_\rms}
\newcommand{\Rc}{R_\rmc}
\newcommand{\Rcg}{\ul{R}_\rmc}
\newcommand{\rhoplg}{\ul{\rho}_{\rmp\rml}}

\newcommand{\alpharc}{\alpha_{r_c}}
\newcommand{\alpharcg}{\ul{\alpha}_{r_{\cg}}}

\newcommand{\taumax}{\tau_{\max}}
\newcommand{\taumaxg}{\ul{\tau}_{\max}}
\newcommand{\tauc}{\tau_c}
\newcommand{\taucg}{\ul{\tau}_{\cg}}
\newcommand{\taurc}{\tau_{r_c}}
\newcommand{\taurcg}{\ul{\tau}_{r_{\cg}}}

\newcommand{\thetagg}{\ul{\theta}}
\newcommand{\thetac}{\theta_c}
\newcommand{\thetacg}{\ul{\theta}_{\cg}}
\newcommand{\varthetarc}{\vartheta_{r_c}}
\newcommand{\varthetarcg}{\ul{\vartheta}_{r_{\cg}}}

\newcommand{\phic}{\phi_c}
\newcommand{\phig}{\ul{\phi}}
\newcommand{\phicg}{\ul{\phi}_{\cg}}
\newcommand{\varphirc}{\varphi_{r_c}}
\newcommand{\varphircg}{\ul{\varphi}_{r_{\cg}}}

\newcommand{\varthetag}{\ul{\vartheta}}
\newcommand{\varphig}{\ul{\varphi}}
\newcommand{\Reff}{R_{\rm{eff}}}
\newcommand{\Tc}{T_{\rmc}}
\newcommand{\dg}{\ul{d}}

\newcommand{\omegag}{\ul{\omega}}
\begin{document}
\bstctlcite{IEEEmax3beforeetal}
\title{Millimeter Wave Beam-Selection Using Out-of-Band Spatial Information}
\author{Anum Ali, {\it Student Member, IEEE}, Nuria Gonz\'alez-Prelcic, {\it Member, IEEE}, and \\Robert W. Heath Jr., {\it Fellow, IEEE}
\thanks{This research was partially supported by the U.S. Department of Transportation through the Data-Supported Transportation Operations and Planning (D-STOP) Tier 1 University Transportation Center and by the Texas Department of Transportation under Project 0-6877 entitled ``Communications and Radar-Supported Transportation Operations and Planning (CAR-STOP)''. N. Gonz\'alez-Prelcic was supported by the Spanish Government and the European Regional Development Fund (ERDF) under Project MYRADA (TEC2016-75103-C2-2-R).}
\thanks{A. Ali and R. W. Heath Jr. are with the Department of Electrical and Computer Engineering, the University of Texas at Austin, Austin, TX 78712-1687 \mbox{(e-mail: \{anumali,rheath\}@utexas.edu)}.}
\thanks{N. Gonz\'alez-Prelcic is with the Signal Theory and Communications Department, University of Vigo, Vigo, Spain \mbox{(e-mail: nuria@gts.uvigo.es)}.}
\thanks{A preliminary version of this work will appear in the Proceedings of
IEEE International Conference on Acoustics, Speech and Signal Processing (ICASSP), New Orleans, USA, March, 2017~\cite{Ali2017Compressed}. }
}
\maketitle
\newtheorem{thm}{Theorem}
\begin{abstract}
Millimeter wave (mmWave) communication is one feasible solution for high data-rate applications like vehicular-to-everything communication and next generation cellular communication. Configuring mmWave links, which can be done through channel estimation or beam-selection, however, is a source of significant overhead. In this paper, we propose to use spatial information extracted at sub-\SI{6}{\giga\hertz}~to help establish the mmWave link. First, we review the prior work on frequency dependent channel behavior and outline a simulation strategy to generate multi-band frequency dependent channels. Second, assuming: (i) narrowband channels and a fully digital architecture at sub-\SI{6}{\giga\hertz}; and (ii) wideband frequency selective channels, OFDM signaling, and an analog architecture at mmWave, we outline strategies to incorporate sub-\SI{6}{\giga\hertz} spatial information in mmWave compressed beam-selection. We formulate compressed beam-selection as a \emph{weighted} sparse signal recovery problem, and obtain the weighting information from sub-\SI{6}{\giga\hertz} channels. In addition, we outline a structured precoder/combiner design to tailor the training to out-of-band information. We also extend the proposed out-of-band aided compressed beam-selection approach to leverage information from all active OFDM subcarriers. The simulation results for achievable rate show that out-of-band aided beam-selection can reduce the training overhead of in-band only beam-selection by $4$x.
\end{abstract}
\clearpage
\section{Introduction}\label{sec:intro}
%
\IEEEPARstart{M}{illimeter wave} (mmWave) communication systems use large antenna arrays and directional beamforming/precoding to provide sufficient link margin~\cite{Pi2011introduction,Rappaport2013Millimeter}. Large arrays are feasible at mmWave as antennas can be packed into small form factors
. Configuring these arrays, however, is not without challenges. First, the high power consumption of RF components makes fully digital baseband precoding difficult~\cite{Pi2011introduction}. Second, the precoder design usually relies on channel state information, which is difficult to acquire at mmWave due to large antenna arrays and low pre-beamforming signal-to-noise ratio (\SNR). Thus, mmWave link establishment has received considerable research interest~\cite{Alkhateeb2014Channel,Choi2015Beam,seo2016training,Wang2009Beam,Hur2013Millimeter}. The usual strategy is to exploit some sort of structure in the unknown channel that aids in link establishment, e.g., sparsity~\cite{Alkhateeb2014Channel,Choi2015Beam} or channel dynamics~\cite{seo2016training}.

Beyond leveraging the structure in the mmWave channel, it is possible to exploit out-of-band (OOB) information extracted from \subsGHz~channels for mmWave link establishment. This is relevant as mmWave systems will likely be deployed in conjunction with lower frequency systems: (i) to provide wide area control signals; and/or (ii) for multi-band communications~\cite{kishiyama2013future,daniels2007multi}. 
In this work, we propose to use the \subsGHz~spatial information as OOB side information about the mmWave channel. This is feasible as the spatial characteristics of \subsGHz~and mmWave channels are similar~\cite{Peter2016Measurement}. Extracting spatial information from \subsGHz~is also enticing due to favorable receive \SNR~in comparison with mmWave. The benefit of using \subsGHz~information in mmWave link establishment has been demonstrated through measurements in~\cite{Nitsche2015Steering}.

OOB information has the potential to reduce the training overhead of mmWave link establishment (be it through channel estimation~\cite{Alkhateeb2014Channel} or beam-selection~\cite{Choi2015Beam}). As such, using OOB information can have a positive impact on all mmWave communication applications. It is, however, especially interesting in high mobility scenarios where frequent link reconfiguration is required. Here, we highlight the benefits in two specific use cases i.e., vehicle-to-everything (V2X) communication and mmWave cellular.

To increase driving automation, next generation vehicles will boast more and better sensors. The sensing ability of a vehicle can be supplemented by exchanging sensor data with other vehicles and infrastructure. Such exchange, however, is data-rate hungry, as sensors may generate up to hundreds of \SI{}{\mega\bitspersecond}~\cite{Angelica2013Googles}. The current vehicular communication mechanisms do not support such data-rates. For example, dedicated short-range communication (DSRC)~\cite{li2010overview} operates at 2-6 \SI{}{\mega\bitspersecond}, and practical rates of LTE-A are limited to several \SI{}{\mega\bitspersecond}~\cite{rumney2013lte}. An amendment of the mmWave WLAN standard IEEE 802.11ad could provide a future framework for a high data-rate V2X communication system, in a similar way as IEEE 802.11p was considered for DSRC.
There is also a growing body of research on mmWave vehicular communications~(see e.g.,~\cite{choi2016millimeter,va2015impact} and the references therein). The highly dynamic nature of V2X channels, however, requires frequent link configuration and OOB-aided link establishment can play an important role in unlocking the potential of mmWave V2X. As an example, the OOB information for mmWave V2X links could come from the DSRC channels or automotive sensors.

Due to the large bandwidths available in the mmWave spectrum and the directional nature of mmWave communications, implying less interference and high data-rate gains, mmWave frequencies are also promising for future outdoor cellular systems~\cite{Andrews2014What,Pi2011introduction,Bai2015Coverage}. Channel measurements have confirmed that mmWave is feasible for both access and backhaul links~\cite{Hur2013Millimeter,Rappaport2013Millimeter}. In addition, the system level evaluation of mmWave network performance has indicated that mmWave cellular systems achieve a spectral efficiency similar to \subsGHz~\cite{Singh2015Tractable}. Understanding these benefits, the Federal Communications Commission (FCC) has proposed rules to make mmWave spectrum available for mobile cellular services~\cite{2016Federal}. At large distances, e.g., cell edges, the pre-beamforming SNR is very low and establishing a reliable mmWave link is particularly challenging. As \SNR~for \subsGHz~sytems is more favorable, reliable OOB information from \subsGHz~can be used to aid the mmWave link establishment.

In this work, we use the \subsGHz~\emph{spatial information} for mmWave link establishment. Specifically, we consider the problem of finding the optimal transmit/receive beam-pair for analog mmWave systems. We assume wideband frequency selective MIMO channels and OFDM signaling for the analog mmWave system. For \subsGHz, we assume narrowband MIMO channels and a fully digital architecture. Both \subsGHz~and mmWave systems use uniform linear arrays (ULAs) at the transmitter (TX) and the receiver (RX). The main contributions of this work are:
\begin{itemize}
\item Based on the review of prior work, we draw conclusions about the expected degree of congruence between \subsGHz~and mmWave systems. The term ``\emph{spatial congruence}'' is used to describe the similarity in the power distribution of arrival/departure angles of the channels. We outline a simulation strategy to generate multi-band frequency dependent channels that are consistent with frequency-dependent channel behavior observed in prior work.
\item We propose a procedure to leverage OOB spatial information in compressed beam-selection~\cite{Choi2015Beam} for an analog mmWave architecture. Specifically, exploiting the limited scattering nature of mmWave channels and using the training on one OFDM subcarrier, we formulate the beam-selection as a weighted sparse signal recovery problem~\cite{Scarlett2013Compressed}. The weights are chosen based on the OOB information. Further, we suggest a structured random codebook design. The proposed design enforces the training precoder/combiner patterns to have high gains in the strong channel directions based on OOB information.
\item We formulate the compressed beam-selection as a multiple measurement vector (MMV) sparse recovery problem~\cite{Tropp2005Simultaneous} to leverage training from all active subcarriers. The MMV based sparse recovery improves the beam-selection by a simultaneous recovery of multiple sparse signals with common support. We extend the weighted sparse recovery and structured codebook design to the MMV case.
\end{itemize}

Prior work on using OOB information in communication systems primarily targets beamforming reciprocity in frequency division duplex (FDD) systems. Based on the observation that the spatial information in the uplink (UL) and downlink (DL) is congruent~\cite{aste1998downlink,HuglSpatial}, several correlation translation strategies were proposed to estimate DL correlation from UL measurements (see~e.g.,~\cite{Jordan2009Conversion,Decurninge2015Channel} and references therein). The estimated correlation was in turn used for DL beamforming. Along similar lines, in~\cite{Vasisht2016Eliminating} the multi-paths in the UL channel were estimated and subsequently the DL channel was constructed using the estimated multi-paths. In~\cite{Shen2016Compressed}, the UL measurements were used as partial support information in compressed sensing based DL channel estimation. The frequency separation between UL and DL is typically small. As an example, there is $9.82\%$ frequency separation between \SI{1935}{\MHz} UL and \SI{2125}{\MHz} DL~\cite{HuglSpatial}. The aforementioned strategies were tailored for the case when the percent frequency separation of the channels under consideration is small and spatial information is congruent. In this work, we consider channels that can have frequency separation of several hundred percent, and hence some degree of spatial disagreement is expected.

There is some prior work on leveraging OOB information for mmWave communications. In~\cite{Nitsche2015Steering}, the directional information from legacy WiFi was used to reduce the beam-steering overhead of \SI{60}{\giga\hertz}~WiFi. The measurement results presented in~\cite{Nitsche2015Steering} confirm the value of OOB information for mmWave link establishment. This work is distinguished from~\cite{Nitsche2015Steering} as the techniques developed in this work are applicable to non line-of-sight (NLOS) channels, whereas~\cite{Nitsche2015Steering} primarily considered LOS channels. In~\cite{ali2016estimating}, correlation translation strategies were presented to estimate mmWave correlation using \subsGHz~correlation. The estimated correlation was in turn used for linear channel estimation in SIMO systems. In contrast with~\cite{ali2016estimating}, we consider MIMO systems and focus on compressed beam-selection in an analog architecture. The concept of radar aided mmWave communication was introduced in~\cite{Gonzalez-Prelcic2016Radar}. The information extracted from a mmWave radar was used to configure the mmWave communication link. Unlike~\cite{Gonzalez-Prelcic2016Radar}, we use \subsGHz~communication system's information for mmWave link establishment.

This work is an extension of~\cite{Ali2017Compressed} and provides a detailed treatment of the problem. Herein, we propose and use a frequency dependent clustered channel model, where each cluster contributes multiple rays. The clustered behavior is observed in recent mmWave channel modeling studies~\cite{Samimi20163}. In contrast,~\cite{Ali2017Compressed} assumed that each cluster contributes a single ray. Further, this work considers frequency selective wideband channels for mmWave, whereas \cite{Ali2017Compressed} assumed a narrowband channel. Finally, we present a structured random codebook design that is not present in~\cite{Ali2017Compressed}.

The rest of the paper is organized as follows: In Section~\ref{sec:multi-bandch}, we review the prior work on frequency dependent channel behavior and outline a simulation strategy to generate multi-band frequency dependent channels. In Section~\ref{sec:sys_ch_model}, we provide the system and channel models for \subsGHz~and mmWave systems. In Section~\ref{sec:CBT}, we outline strategies to incorporate the OOB information in compressed beam-selection. The simulation results are presented in Section~\ref{sec:simres}, and Section~\ref{sec:conc} concludes the paper.

\emph{Notation:} We use the following notation throughout the paper. Bold lowercase $\bx$ is used for column vectors, bold uppercase $\bX$ is used for matrices, non-bold letters $x$, $X$ are used for scalars. $[\bx]_m$, $[\bX]_{m,n}$, $[\bX]_{m,:}$, and $[\bX]_{:,n}$, denote $m$th entry of $\bx$, entry at the $m$th row and $n$th column of $\bX$, $m$th row of $\bX$, and $n$th column of $\bX$, respectively. Superscript $\transp$ and $\ast$ represent the transpose and conjugate transpose. $\bzero$ and $\bI$ denote the zero vector and identity matrix respectively. $\cC\cN(\bx,\bX)$ denotes a complex circularly symmetric Gaussian random vector with mean $\bx$ and correlation $\bX$. We use $\bbE[\cdot]$, $\|\!\cdot\!\|_p$, and $\|\!\cdot\!\|_\rmF$ to denote expectation, $p$ norm and Frobenius norm, respectively. $\bX\otimes\bY$ is the Kronecker product of $\bX$ and $\bY$. Calligraphic letter $\cX$ denotes sets and $[X]$ represents the set $\{1,2,\cdots,X\}$. Finally, $|\!\cdot\!|$ is the absolute value of its argument or the cardinality of a set, and $\vec(\cdot)$ yields a vector for a matrix argument. The \subsGHz~variables are underlined, as $\ul{\bx}$, to distinguish them from mmWave.
\section{Multi-band channel characteristics and simulation}\label{sec:multi-bandch}
The OOB-aided mmWave beam-selection strategies proposed in this work rely on the information extracted at \subsGHz. Therefore, it is essential to understand the similarities and differences between \subsGHz~and mmWave channels. Furthermore, to assess the performance of proposed OOB-aided mmWave link establishment strategies, a simulation strategy is required to generate multi-band frequency dependent channels. In this section, we review a representative subset of prior work to draw conclusions about the expected degree of spatial congruence between \subsGHz~and mmWave channels. Based on these results, we outline a strategy to simulate multi-band frequency dependent channels.

Due to the differences in the wavelength of \subsGHz~and mmWave frequencies, it is possible that the Fresnel zone clarity criterion for LOS is satisfied at \subsGHz~but not for mmWave. It is expected that the OOB-aided link establishment will not perform well in such scenarios as the spatial information in a LOS \subsGHz~and NLOS mmWave may be different. One can venture to detect such scenarios and revert to in-band only link establishment. The authors defer this issue to future work.
\subsection{Review of multi-band channel characteristics}\label{sec:chbh}
The material properties change with frequency, e.g., the relative conductivity and the average reflection increase with frequency~\cite{ITU2001Propagation,Violette1988Millimeter}. Hence, some characteristics of the channel are expected to vary with frequency. It was reported that the delay spread decreases~\cite{weiler2015simultaneous,Poon2003Indoor,Jaeckel20165G,Kaya201628}, the number of angle-of-arrival (AoA) clusters increase~\cite{Samimi20163}, the shadow fading increases~\cite{Jaeckel20165G}, and the angle spread (AS) of clusters decreases~\cite{Kaya201628,Poon2003Indoor} with frequency. Further, it was observed that the late arriving multi-paths have more frequency dependence due to higher interactions with the environment~\cite{Qiu1999Multipath,haneda2012modeling}.

Not all channel characteristics vary greatly with frequency. As an example, the existence of spatial congruence between the UL and DL channels is well established~\cite{aste1998downlink,HuglSpatial}. In~\cite{aste1998downlink}, it was noted that though the propagation channels in UL and DL are not reciprocal, the spatial information is congruent. It was observed in measurements (for \SI{1935}{\mega\hertz} UL and \SI{2125}{\mega\hertz} DL) that the deviation in AoAs of dominant paths of UL and DL is small with high probability~\cite{HuglSpatial}. Prior work has exploited the spatial congruence between UL and DL channels to reduce/eliminate the feedback in FDD systems, see e.g., \cite{Jordan2009Conversion,Decurninge2015Channel,Vasisht2016Eliminating,Shen2016Compressed}.

Some channel characteristics are congruent for larger frequency separations. In~\cite{Peter2016Measurement}, the directional power distribution of $5.8$ \GHz, $14.8$ \GHz, and $58.7$ \GHz~LOS channels were reported to be almost identical. The number of resolvable paths, the decay constants of the clusters, the decay constants of the subpaths within the clusters, and the number of angle-of-departure (AoD) clusters were found to be similar in $28$ and $73$ \GHz~channels~\cite{Samimi20163}. In~\cite{Dupleich2016Simultaneous}, similar power delay profiles (PDP)s were reported for $10$ \GHz~and $30$ \GHz~channels. The received power as a function of distance was found to be similar for $5.8$ \GHz~and $14.8$ \GHz~in~\cite{Peter2016Measurement}. Only minor differences were observed in the CDFs of delay spread, azimuth AoA/AoD spread, and elevation AoA/AoD spread of six different frequencies between $2~\GHz$ and $60$ \GHz~in \cite{Ky2016Frequency}.

In conclusion, the prior work confirms that there is substantial similarity between channels at different frequencies, even with large separations. Hence, it is likely that there is significant, albeit not perfect, congruence between \subsGHz~and mmWave channels. This observation is leveraged by prior work that used legacy WiFi measurements to configure $60~\GHz$ WiFi links~\cite{Nitsche2015Steering}.

\subsection{Simulation of multi-band frequency dependent channels}\label{subsec:ch_gen_algo}
The following observations are made about the frequency dependent channel behavior from the review of the prior work:
\begin{itemize}
\item The channel characteristics differ with frequency, and the differences increase as the percent separation between center frequencies of the channels increase.
\item The late arriving multi-paths have more frequency dependence~\cite{Qiu1999Multipath,haneda2012modeling}, and some paths may be present at one frequency but not at the other~\cite{Haneda2007Experimental}.
\end{itemize}

The proposed multi-band frequency dependent channel simulation algorithm takes the aforementioned observations into consideration. It takes the parameters of the channels at two frequencies as input and outputs a random realization for each of the two channels. The input parameters include the number of clusters, the number of paths within a cluster, maximum delay spread, root mean squared (RMS) time spread of the paths within clusters, center frequency, and the RMS AS of the paths within clusters. The output random realizations of the two channels are consistent in the sense that one of the channels is a perturbed version of the other, where the perturbation model respects the frequency dependent channel behavior. Before discussing the proposed simulation algorithm, we present the required preliminaries.

The following exposition is applicable to the channels at two frequencies $f_1$ and $f_2$ (not necessarily \subsGHz~and mmWave). Therefore, we use subscript index $i\in[I]$, where $I=2$, to distinguish the parameters of the channel at center frequency $f_1$ from the parameters of the channel at center frequency $f_2$. In subsequent sections, we use underlined and non-underlined analogs of the variables defined here for \subsGHz~and mmWave, respectively. We assume that there are $C_i$ clusters in the channel $i$. Each cluster has a mean time delay $\tau_{c,i}\in [0,\tau_{\max,i}]$ and mean physical AoA/AoD $\{\theta_{c,i}, \phi_{c,i}\} \in [0,2\pi)$. Each cluster $c_i$ is further assumed to contribute $R_{c,i}$ rays/paths between the TX and the RX. Each ray $r_{c,i}\in[R_{c,i}]$ has a relative time delay $\tau_{r_{c,i}}$, relative AoA/AoD shift $\{\vartheta_{r_{c,i}},\varphi_{r_{c,i}}\}$, and complex path gain $\alpha_{r_{c,i}}$. If $\rho_{\rmp\rml,i}$ represents the path-loss, then the omni-directional impulse response of the channel $i$ can be written as
\begin{align}
h_{\rmo\rmm\rmn\rmi,i}(t,\theta,\phi)=\dfrac{1}{\sqrt{\rho_{\rmp\rml,i}}} \sum_{c_i=1}^{C_i} \sum_{r_{c,i}=1}^{R_{c,i}} \alpha_{r_{c,i}} \delta(t-\tau_{c,i}-\tau_{r_{c,i}})\times
\delta(\theta-\theta_{c,i}-\vartheta_{r_{c,i}})
\times\delta(\phi-\phi_{c,i}-\varphi_{r_{c,i}}).
\label{eq:homni}
\end{align}
The continuous time channel impulse response given in~\eqref{eq:homni} is not band-limited. The impulse response convolved with pulse shaping filter, however, is band-limited and can be sampled to obtain the discrete time channel as in Section~\ref{sec:sys_ch_model}. Further, in~\eqref{eq:homni} we have only considered the azimuth AoAs/AoDs for simplicity. The general formulation with both azimuth and elevation angles is a straightforward extension, see~\cite{Samimi20163}. A detailed discussion on the choice of the channel parameters is beyond the scope of this paper. The reader is directed to prior work e.g.,~\cite{Nurmela2015METIS} for discussions on suitable channel parameters. A cursory guideline can be established, however, based on the literature review presented earlier. Assuming $f_1\leq f_2$ it is expected that $C_1\geq C_2$~\cite{Samimi20163}, $R_{c,1} \geq R_{c,2}$~\cite{Samimi20163}, $\tau_{\max,1} \geq \tau_{\max,2}$~\cite{weiler2015simultaneous,Poon2003Indoor,Jaeckel20165G,Kaya201628}, $\sigma_{\vartheta_{c,1}} \geq \sigma_{\vartheta_{c,2}}$, and $\sigma_{\varphi_{c,1}} \geq \sigma_{\varphi_{c,2}}$~\cite{Kaya201628,Poon2003Indoor}. Furthermore, the parameters used for numerical evaluations in Section~\ref{sec:simres} are an example of the parameters that comply with the observations of the prior work.

The proposed channel simulation strategy is based on a two-stage algorithm. In the first stage, the mean time delays $\tau_{c,i}$ and mean AoAs/AoDs $\{\theta_{c,i},\phi_{c,i}\}$ of the clusters are generated for both frequencies in a coupled manner, while respecting the frequency dependent behavior. In the second stage, the paths within the clusters are generated independently for both frequencies. The first stage of the proposed channel simulation algorithm is outlined in Algorithm~\ref{alg:first_stage}.

\subsubsection{First Stage}
The number of clusters $C_i$, the maximum delay spreads $\tau_{\max,i}$, and the center frequencies $f_i$ for channels $i\in[I]$ are fed to the first-stage of the proposed algorithm as inputs. The first stage has three parts. In the first part, the clusters for both the channels are generated independently. In the second part, we replace several clusters in one of the channels by the clusters of the other channel. The first two parts ensure that there are a few common as well as a few uncommon clusters in the channels. Finally, in the third part frequency dependent perturbations are added to the clusters of one of the channels. This is to imitate the effect that the common clusters in the two channels may have a time/angle offset.

\textbf{Part 1: (Generation)} The algorithm initially generates the mean time delays and mean AoAs/AoDs for the clusters i.e., $\{\tau_{c,i},\theta_{c,i},\phi_{c,i}\}$, $\forall c_i\in[C_i], \forall i\in[I]$. The set of the three parameters $\{\tau_{c,i},\theta_{c,i},\phi_{c,i}\}$ corresponding to a cluster is referred to as the cluster parameter set. The clusters for both channels are generated independently.

\textbf{Part 2: (Replacement)} We replace several clustes in one channel with the clusters of the other to esnure common clusters in the channels. This step is to be carried in accordance with the following observations: (i) the late arriving clustered paths are more likely to fade independently across the two channels~\cite{Qiu1999Multipath,haneda2012modeling}; and (ii) independent clustered paths are more likely as the percent frequency separation increases. In other words, for fixed percent frequency separation, the early arriving clustered paths have a higher likelihood of co-occurrence in both channels. We store the indices of co-occuring clusters in sets $\cR_i,\forall i\in[I]$, henceforth called the replacement index sets. As an example, the index sets can be created by sorting the cluster parameter sets in an ascending order with respect to $\tau_{c,i}$ and populating $\cR_i =\{ i : \xi> \frac{|f_i-f_{[[I]\setminus i]}|}{\max(f_i,f[[I]\setminus i])} \frac{\tau_{c,i}}{\tau_{\max,i}}\}$. Here $\xi$ is a standard Uniform random variable, i.e., $\xi\sim U[0,1]$. For the candidate indices that appear in $\cR_1$ and $\cR_2$, we replace the corresponding clusters in one channel with those of the other. Without loss of generality, we replace the clusters of the channel with larger delay spread. Hence, we update the cluster parameters sets $\{\tau_{c,b},\theta_{c,b},\phi_{c,b}\}~\forall c_b \in \cR_b\cap\cR_{[I] \setminus b}$ with $\{\tau_{c,[I]\setminus b},\theta_{c,[I] \setminus b},\phi_{c, [I] \setminus b}\}~\forall c_{[I]\setminus b} \in \cR_b\cap\cR_{[I] \setminus b}$, where $b=\underset{i}{\arg\max} ~\tau_{\max,i}$.

\textbf{Part 3: (Perturbation)} So far we have simulated the effect that there will be common as well as uncommon clusters in the channels at two frequencies. Now we need to add frequency dependent perturbation in the clusters of one of the channels to simulate the behavior that co-occurring clusters can have some time/angle offset. The perturbation should be: (i) proportional to the mean time delay of the cluster~\cite{Qiu1999Multipath,haneda2012modeling}; and (ii) proportional to percent center frequency separation. We continue by assuming that the clusters of the channel $b$ are perturbed. A scalar perturbation $\Delta_{c,b}$ is generated $\forall c_b\in [C_b]$. The perturbation $\Delta_{c,b}$ is then modified for delays and AoAs/AoDs using deterministic modifiers $g_{\tau}(\cdot)$, $g_{\theta}(\cdot)$ and $g_{\phi}(\cdot)$, respectively. The rationale of using deterministic modifications of the same perturbation for delays and angles is the coupling of these parameters in the physical channels. This is to say that the amount of variation in the mean delay of the cluster, from one frequency to another, is not expected to be independent of the variation in AoA/AoD. Let us define the function
\begin{align}
q(x,w,y,z)=\begin{cases}
1&\text{if }x-w<y,\\
-1&\text{if }x+w>z,\\
\pm 1~\text{with equal probability}&\text{otherwise}.\\
\end{cases}
\end{align}
With this definition, an example perturbation model could be $\Delta_{c,b} \sim U[0,1]$, $g_{\tau}(\Delta_{c,b})=q(\tau_{c,b},\Delta_{c,b},0,\tau_{\max,b})\frac{|f_b-f_{[I]\setminus b}|}{\max(f_b,f_{[I]\setminus b})}\tau_{c,b}\Delta_{c,b}$ and $g_{\theta}(\Delta_{c,b})=q(\theta_{c,b},\Delta_{c,b},0,2\pi)\frac{|f_b-f_{[I]\setminus b}|}{\max(f_b,f_{[I]\setminus b})}\frac{\tau_{c,b}}{\tau_{\max,b}}\Delta_{c,b}$. The modifier $g_{\phi}$ can be chosen similar to $g_{\theta}$. The modified perturbations $g_{\tau}(\Delta_{c,b})$, $g_{\theta}(\Delta_{c,b})$, and $g_{\phi}(\Delta_{c,b})$ are added in $\tau_{c,b}$, $\theta_{c,b}$, and $\phi_{c,b}$, respectively, to obtain the cluster parameters for channel $b$. Finally, the cluster parameter sets for both channels are returned.

\begin{algorithm}
\caption{Mean time delays $\tau_{c,i}$ and mean AoAs/AoDs $\{\theta_{c,i},\phi_{c,i}\}$ generation}
\begin{algorithmic}[1]
\INPUT $C_i,\tau_{\max,i},f_i,~\forall i\in[I]$
\OUTPUT $\{\tau_{c,i},\theta_{c,i},\phi_{c,i}\}~\forall c_i\in[C_i], \forall i\in [I]$
\STATE Draw $\tau_{c,i}\sim U[0,\tau_{\max,i}], \{\theta_{c,i},\phi_{c,i} \} \sim U [0,2\pi) ~ \forall c_i \in [C_i], \forall i\in[I]$ to get the cluster parameter sets $\{\tau_{c,i},\theta_{c,i},\phi_{c,i}\}~\forall c_i\in[C_i], \forall i\in[I]$.
\STATE Populate replacement index sets $\cR_i,\forall i\in[I]$ using a suitable replacement model, and update $\{\tau_{c,b},\theta_{c,b},\phi_{c,b}\}~\forall c_b \in \cR_b\cap\cR_{[I] \setminus b } \leftarrow \{\tau_{c,[I]\setminus b},\theta_{c,[I] \setminus b},\phi_{c, [I] \setminus b}\}~\forall c_{[I]\setminus b} \in \cR_b\cap\cR_{[I] \setminus b }$, where $b=\underset{i}{\arg\max}~\tau_{\max,i}$.
\STATE Generate $C_b$ perturbations $\Delta_{c,b}$ and update $\tau_{c,b}\leftarrow\tau_{c,b}+g_{\tau}(\Delta_{c,b})$, $\theta_{c,b}\leftarrow\theta_{c,b}+g_{\theta}(\Delta_{c,b})$, and $\phi_{c,b}\leftarrow \phi_{c,b}+g_{\phi}(\Delta_{c,b})$.
\end{algorithmic}
\label{alg:first_stage}
\end{algorithm}

\subsubsection{Second Stage}
Once the parameters $\{\tau_{c,i},\theta_{c,i},\phi_{c,i}\},\forall c_i\in[C_i],\forall i\in[I]$ are available, the paths/rays within the clusters are generated independently for both channels in the second stage of the proposed algorithm. The second stage requires the number of paths per cluster $R_{c,i}$, the RMS time spread of the paths within clusters $\sigma_{\tau_{r_{c,i}}}$, and the RMS AS of the relative arrival and departure angle offsets $\{\sigma_{\vartheta_{c,i}},\sigma_{\varphi_{c,i}}\}$, as input. The output of the second stage are the sets $\{\alpha_{r_{c,i}},\tau_{r_{c,i}},\vartheta_{r_{c,i}},\varphi_{r_{c,i}}\},~\forall r_{c,i} \in R_{c,i},\forall i\in[I]$. The relative time delays $\tau_{r_{c,i}}$ are generated according to a suitable intra-cluster PDP (e.g., Exponential or Uniform), the relative angle shifts $\{\vartheta_{r_{c,i}},\varphi_{r_{c,i}}\}$ according to a suitable azimuth power spectrum (APS) (e.g., uniform, truncated Gaussian or truncated Laplacian), and the complex coefficients  $\alpha_{r_{c,i}}$ according to a suitable fading model (e.g., Rayleigh or Ricean).

\section{System and channel model}\label{sec:sys_ch_model}
\begin{figure}[h!]
\centering
\includegraphics[width=0.45\textwidth]{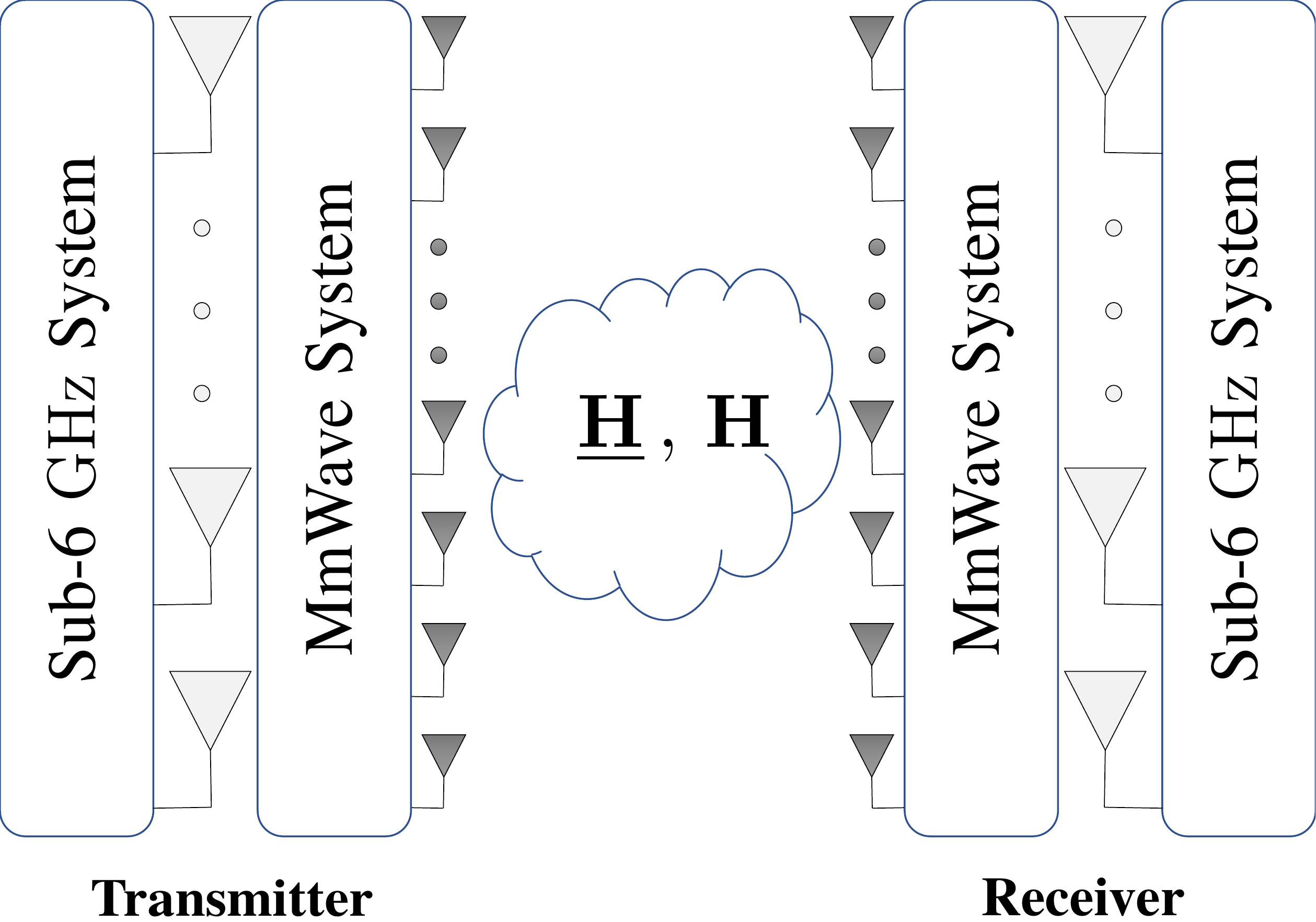}
\caption{The multi-band MIMO system with co-located \subsGHz~and mmWave antenna arrays. The \subsGHz~channel is $\bHg$ and the mmWave channel is $\bH$.}
\label{fig:TXRX}
\end{figure}

We consider a multi-band MIMO system shown in Fig.~\ref{fig:TXRX}, where ULAs of isotropic point sources are used at the TX and the RX. The ULAs are considered for ease of exposition, whereas, the proposed strategies can be extended to other array geometries with suitable modifications. We assume that the \subsGHz~and mmWave arrays are co-located, aligned, and have comparable apertures. Both \subsGHz~and mmWave systems operate simultaneously.
\subsection{\SubsGHz~system and channel model}\label{subsec:sub6GHz_sys_ch_model}
\begin{figure}[h!]
\centering
\includegraphics[width=0.5\textwidth]{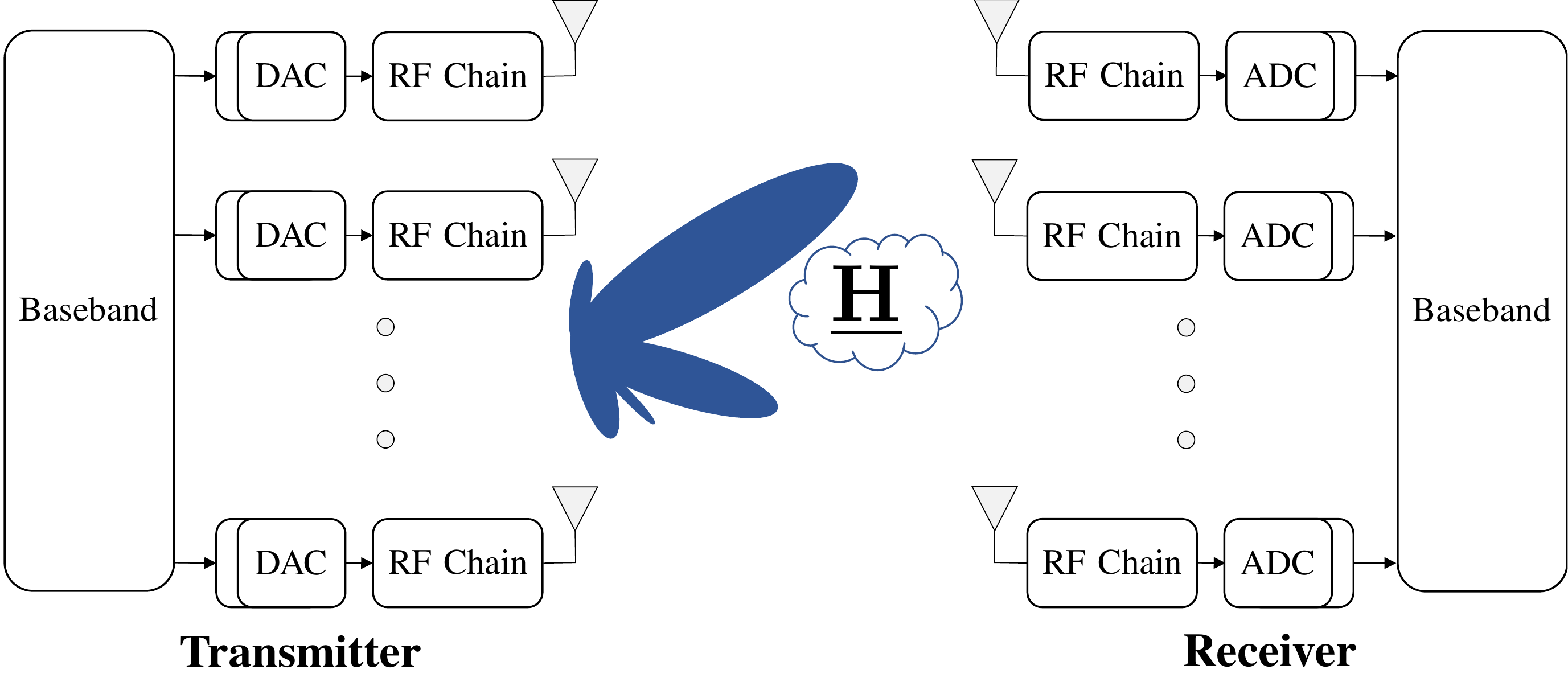}
\caption{The \subsGHz~system with digital precoding.}
\label{fig:sub6GHzsystem}
\end{figure}

The \subsGHz~system is shown in Fig.~\ref{fig:sub6GHzsystem}. Note that, we underline all \subsGHz~variables to distinguish them from the mmWave variables. The \subsGHz~system has one RF chain per antenna and as such, fully digital precoding is possible. 
We assume narrowband signaling at \subsGHz. Extending the proposed approach to wideband \subsGHz~systems is straight forward because only the directional information is retrieved from \subsGHz, which is not expected to vary much across the channel bandwidth. 
We adopt a geometric channel model for $\bHg$ based on~\eqref{eq:homni}. The MIMO channel matrix for \subsGHz~can be written as
\begin{align}
\bHg=\sqrt{\dfrac{\MRXg\MTXg}{\rhoplg}} \sum_{\cg=1}^{\Cg} \sum_{r_{\cg}=1}^{\Rcg} \alpharcg p (-\taucg - \taurcg) \aRXg(\thetacg+\varthetarcg) \aTXg^\ast(\phicg+\varphircg),
\label{eq:subsghzchmodel}
\end{align}
where $p(\tau)$ denotes a pulse shaping function evaluated at $\tau$ seconds, $\aRXg(\theta)$ and $\aTXg(\phi)$ are the antenna array response vectors of the RX and the TX, respectively. The array response vector of the RX is
\begin{align}
\aRXg(\theta)=\frac{1}{\sqrt{\MRXg}}[1,e^{\compj 2\pi\dg\sin(\theta)},\cdots,e^{\compj 2\pi(\MRXg-1)\dg\sin(\theta)}]^\transp,
\end{align}
where $\dg$ is the inter-element spacing in wavelength. The array response vector of the TX is defined in a similar manner. The normalized spatial AoA of the \subsGHz~system is $\ul{\nu}\triangleq\ul{d}\sin(\theta)$ and the spatial AoD is $\ul{\omega}\triangleq\ul{d}\sin(\phi)$. 

\subsection{Millimeter wave system and channel model}\label{subsec:mmWave_sys_ch_model}
\begin{figure}[h!]
\centering
\includegraphics[width=0.5\textwidth]{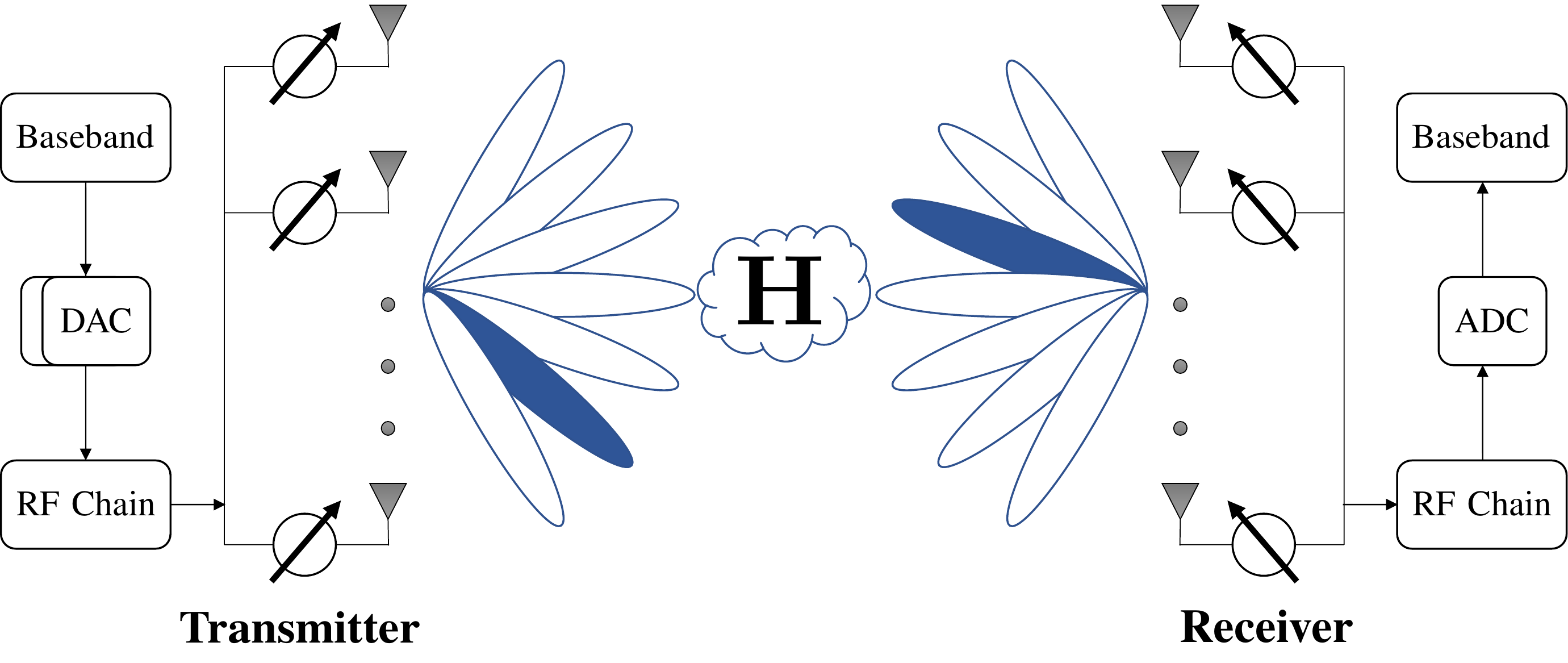}
\caption{The mmWave system with phase-shifters based analog beamforming.}
\label{fig:mmWavesystem}
\end{figure}

The mmWave system is shown in Fig.~\ref{fig:mmWavesystem}. The TX has $\MTX$ antennas and the RX has $\MRX$ antennas. Both the TX and the RX are equipped with a single RF chain, hence only analog beamforming is possible. The idea of using OOB information can also be applied to hybrid analog/digital and fully digital low-resolution mmWave architectures, an interesting direction for future work. The mmWave system uses OFDM signaling with $K$ subcarriers. The data symbols on each subcarrier $\sfs[k],~\forall k\in[K]$ are transformed to the time-domain using a $K$-point IDFT. A cyclic prefix (CP) of length $\Lc$ is then prepended to the time-domain samples before applying the analog precoder $\bff$. The length $\Lc$ CP followed by the $K$ time-domain samples constitute one OFDM block. The effective transmitted signal on subcarrier $k$ is $\bff \sfs[k]$. The data symbols follow $\bbE[\sfs[k]\sfs^\ast[k]]=\dfrac{\Pt}{K}$, where $\Pt$ is the total average power in the useful part, i.e., ignoring the CP, per OFDM block. Since $\bff$ is implemented using analog phase-shifters, it has constant modulus entries i.e., $|[\bff]_m|^2=\frac{1}{\MTX}$. Further, we assume that the angles of the analog phase-shifters are quantized and have a finite set of possible values. With these assumptions, $[\bff]_m=\frac{1}{\sqrt{\MTX}}e^{\compj \zeta_{m}}$, where $\zeta_m$ is the quantized angle.

We assume perfect time and frequency synchronization at the receiver. The received signal is first combined using an analog combiner $\bq$. The CP is then removed and the time-domain samples are converted back to the frequency-domain using a $K$-point DFT. If the $\MRX\times \MTX$ MIMO channel at the subcarrier $k$ is denoted as $\bsfH[k]$, the received signal on subcarrier $[k]$ after processing can be expressed as
\begin{align}
\check\sfy[k]=\bq^\ast \bsfH[k] \bff \sfs[k] + \bq^\ast \check\bsfv[k],
\label{eq:rxd_sgnl}
\end{align}
where $\check\bsfv[k]\sim \cC\cN(\bzero,\noisepsf \bI), \forall k\in[K]$.

Several parameters used in the mmWave channel model are analogous to the \subsGHz~counterparts in~\eqref{eq:subsghzchmodel}. As such, in the sequel we only discuss the parameters that have some distinction from \subsGHz. Due to the wideband nature of mmWave communications, the channel is assumed to be frequency selective. We assume that the channel has $L$ taps where $L\leq\Lc+1$. Under this model, the delay-$\ell$ MIMO channel matrix $\bH[\ell]$ can be written as
\begin{align}
\bH[\ell]=\sqrt{\dfrac{\MRX\MTX}{\rhopl}} \sum_{c=1}^C \sum_{r_c=1}^{R_c} \alpharc p (\ell\Ts-\tauc-\taurc)\aRX(\thetac+\varthetarc)\aTX^\ast(\phic+\varphirc),
\label{eq:timedomch}
\end{align}
where $\Ts$ is the signaling interval. With the delay-$\ell$ MIMO channel matrix given in~\eqref{eq:timedomch}, the channel at subcarrier $k$, $\bsfH[k]$ can be expressed as
\begin{align}
\bsfH[k]=\sum_{\ell=0}^{L-1}\bH[\ell] e^{-\compj \tfrac{2\pi k}{K} \ell}.
\label{eq:freqdomch}
\end{align}

\section{Compressed beam-selection with out-of-band information}\label{sec:CBT}
We begin this section by formulating the compressed beam-selection problem. We initially formulate the problem using training from a single subcarrier. We then proceed to incorporate OOB information in the compressed beam-selection strategy by using a weighted sparse recovery algorithm and structured codebooks. Finally, we use the MMV formulation to extend the proposed OOB-aided compressed beam-selection to leverage training data from all active subcarriers.
\subsection{Problem formulation}
In the training phase, if the TX uses a training precoding vector $\bff_m$ and the RX uses a training combining vector $\bq_n$, then by the mmWave system model \eqref{eq:rxd_sgnl}, the received signal on the $k$th subcarrier is
\begin{align}
\check\sfy_{n,m}[k]=\bq_n^\ast\bsfH[k]\bff_m \sfs_m[k] + \bq_n^\ast \check\bsfv_{n,m}[k],
\end{align}
where $\bff_m\sfs_m[k]$ is the precoded training symbol on subcarrier $k$. The TX transmits the training OFDM blocks on $\NTX$ distinct precoding vectors. For each precoding vector, the RX uses $\NRX$ distinct combining vectors. The number of total training blocks is $\NRX\times \NTX$. For simplicity we assume that the transmitter uses $\sfs_m[k]=\sfs[k]$ throughout the training period. Collecting the received signals and dividing through by the training, we get an $\NRX\times\NTX$ matrix
\begin{align}
\bsfY[k]=\bQ^\ast\bsfH[k]\bF+\bsfV[k],
\label{eq:measurements}
\end{align}
where $\bQ=[\bq_1,\bq_2,\cdots,\bq_{\NRX}]$ is the $\MRX\times \NRX$ combining matrix, $\bF=[\bff_1,\bff_2,\cdots,\bff_{\NTX}]$ is the $\MTX \times \NTX$ precoding matrix, and $\bsfV[k]$ is the $\NRX\times \NTX$ post-processing noise matrix after combining and division by the training. 

For analog beamforming, the phase of the signal transmitted from each antenna is controlled by a network of analog phase-shifters. If $\DTX=\log_2(\MTX)$ bit phase-shifters are used at the TX, and similarly $\DRX=\log_2(\MRX)$, then the DFT codebooks can be realized. If we define $\nu_m=\tfrac{2m-1-\MRX}{2\MRX}$ and $\theta_m=\arcsin(\frac{\nu_m}{d})$, then the $m$th codeword in the codebook for the RX is $\aRX(\theta_m)$. The DFT codebook for the TX is similarly defined using  $\omega_m\triangleq\tfrac{2m-1-\MTX}{2\MTX}$ and $\phi_m=\arcsin(\frac{\omega_m}{d})$. We denote the DFT codebook for the RX as $\ARX$ and for the TX as $\ATX$. If the DFT codebooks are used in the training phase, then~\eqref{eq:measurements} (in the absence of noise) can be written as
\begin{align}
\bsfG[k]=\ARX^\ast\bsfH[k]\ATX.
\label{eq:beamspace}
\end{align}
The matrix $\bsfG[k]$ in \eqref{eq:beamspace} is the \emph{beamspace} or \emph{virtual} representation of the channel $\bsfH[k]$~\cite{Sayeed2002Deconstructing}. From the unitary nature of DFT matrices we can also write $\bsfH[k]=\ARX\bsfG[k]\ATX^\ast$. We give the vectorized form of the channel $\bsfH[k]$ below, that will come in handy in the subsequent developments.
\begin{align}
\vec(\bsfH[k])=(\ATX^\rmc\otimes\ARX) \vec(\bsfG[k])=(\ATX^\rmc\otimes\ARX)\bsfg[k].
\label{eq:vechk}
\end{align}

Due to limited scattering of the mmWave channel, the virtual representation of every channel tap is sparse~\cite{Sayeed2002Deconstructing} if the grid offset related leakage is ignored. If the impact of the grid offset is considered, the virtual matrix for every delay tap is less sparse due to leakage~\cite{Schniter2014Channel}. Moreover, the frequency domain channel matrices $\bsfH[k]$ result by the summation of the MIMO matrices for all the channel taps, see~\eqref{eq:freqdomch}. The sparsity pattern in the virtual representations of different delays is not the same. Hence, the number of active coefficients of the frequency domain MIMO channel at any subcarrier is larger than the number of active coefficients in the virtual representation of a single delay tap MIMO channel. With large enough antenna arrays at the transmitter and receiver and a few clusters with small AS in the mmWave channel, however, the virtual representation of the MIMO channel corresponding to any subcarrier can be considered approximately sparse. As such, we proceed by assuming that $\bsfG[k]$ is a sparse matrix.
In beam-selection, we seek only the largest entry of $\bsfg[k]$. A slight modification of the subsequent formulation, however, can also be used for channel estimation~\cite{Alkhateeb2015Compressed}.
The index of the largest absolute entry in $\bsfg[k]$, i.e., $r^\star=\underset{1\leq r \leq \MRX\MTX}{\arg\max} |[\bsfg[k]]_r|$, determines the best beam-pair (or codewords). Specifically, the best transmit beam index is $j^\star=\lceil \frac{r^\star}{\MRX} \rceil$, and the best receive beam index is $i^\star=r^\star-(j^\star-1)\MRX$. The receiver needs to feedback the best transmit beam index to the transmitter, which can be achieved using the active \subsGHz~link. Note that we did not keep the index $[k]$ with $r$ as the analog transmit and receive beams are independent of the subcarrier.

Reconstructing $\bsfG[k]$ (or $\bsfg[k]$) by exhaustive-search as in~\eqref{eq:beamspace} incurs a training overhead of $\MTX\times\MRX$ blocks. The training burden can be reduced by exploiting the sparsity of $\bsfg[k]$. The resulting framework, called compressed beam-selection, uses a few random measurements of the space to estimate $r^\star$. The training codebooks that randomly sample the space while respecting the analog beamforming constraints were reported in~\cite{Alkhateeb2015Compressed}, where TX designs its $\MTX\times\NTX$ training codebook such that $[\bF]_{n,m}=\frac{1}{\sqrt{\MTX}}e^{\compj \zeta_{n,m}}$, where $\zeta_{n,m}$ is randomly and uniformly selected from the set of quantized angles $\{0,\frac{2\pi}{2^{\DTX}},\cdots,\frac{2\pi(2^{\DTX}-1)}{2^{\DTX}}\}$. The RX similarly designs its $\MRX\times\NRX$ training codebook $\bQ$. To formulate the compressed beam-selection problem \eqref{eq:measurements} is vectorized to get
\begin{align}
\bsfy[k]=\vec(\bsfY[k])&=(\bF^\transp\otimes\bQ^\ast)\vec(\bsfH[k])+\vec(\bsfV[k]),\nonumber\\
&\overset{(a)}{=} (\bF^\transp\otimes\bQ^\ast) (\ATX^\rmc\otimes\ARX) \bsfg[k]\!+\vec(\bsfV[k]),\nonumber\\
&\overset{(b)}{=} \bPsi \bsfg[k] + \vec(\bsfV[k]).
\label{eq:CSProb}
\end{align}
In~\eqref{eq:CSProb}, $(a)$ follows from~\eqref{eq:vechk} and $(b )$ follows by introducing the sensing matrix $\bPsi=(\bF^\transp \otimes \bQ^\ast)(\ATX^\rmc\otimes\ARX)$. Exploiting the sparsity of $\bsfg[k]$, $r^\star$ can be estimated reliably, even when $\NTX \ll \MTX$ and $\NRX \ll \MRX$. The system~\eqref{eq:CSProb} can be solved for sparse $\bsfg[k]$ using any of the sparse signal recovery techniques. In this work, we use the orthogonal matching pursuit (OMP) algorithm~\cite{Tropp2007Signal}. We outline the working principle of OMP here and refer the interested readers to~\cite{Tropp2007Signal} for details. The OMP algorithm uses a greedy approach in which the support is constructed in an incremental manner. At each iteration, the OMP algorithm adds to the support estimate the column of $\bPsi$ that is most highly correlated with the residual. The measurement vector $\bsfy[k]$ is used as the first residual vector, and subsequent residual vectors are calculated as $\tilde\bsfy[k]=\bsfy[k] - \bPsi \hat\bsfg[k]$, where $\hat\bsfg[k]$ is the least squares estimate of $\bsfg[k]$ on the support estimated so far. As we are interested only in $r^\star$, we can find the approximate solution in a single step using the OMP framework, i.e.,
\begin{align}
r^\star=\underset{1\leq r \leq \MRX\MTX}{\arg\max} |[\bPsi]_{:,r}^\ast\bsfy[k]|.
\label{eq:OMPproj}
\end{align}
\subsection{Proposed two-stage OOB-aided compressed beam-selection}
The proposed OOB-aided compressed beam-selection is a two-stage procedure. In the first stage, the spatial information is extracted from \subsGHz~channel. In the second stage, the extracted information is used for compressed beam-selection.
\subsubsection{First stage (spatial information retrieval from \subsGHz)}\label{sec:info_retr}
The spatial information sought from \subsGHz~is the dominant spatial directions i.e., AoAs/AoDs. Prior work has considered the specific problem of estimating both the AoAs/AoDs~(see e.g.,\cite{Bencheikh2010Polynomial}) and the AoAs/AS~(see e.g.,~\cite{bengtsson2000low}) from an empirically estimated spatial correlation matrix. The generalization of these strategies to joint AoA/AoD/AS estimation, however, is not straightforward. Further, for rapidly varying channels, a reliable estimate of the channel correlation is also difficult to acquire. Therefore, we seek a methodology that can provide reliable spatial information for rapidly varying channels with minimal overhead. For the application at hand, the demand on the accuracy of the direction estimates, however, is not particularly high. Due to the inherent differences between \subsGHz~and mmWave channels, the extracted spatial information will have an unavoidable mismatch. Consequently, we only need a coarse estimate of the angular information from \subsGHz.

The earlier work on AoA/AoD estimation was primarily inspired by spectrum estimation. Fourier analysis is perhaps the most basic spectral estimation approach. In this work, we use the spatial Fourier transform of the MIMO channel $\bHg$ (i.e., spatial spectrum) to obtain a coarse estimate of the dominant directions. The MIMO channel $\bHg$ is required for the operation of \subsGHz~system itself. Hence, the direction estimation based on Fourier analysis does not incur any additional training overhead from OOB information retrieval point of view.

\begin{figure}[h!]
\centering
    \centering
    \begin{subfigure}[t]{0.45\textwidth}
        \centering
        \includegraphics[width=\textwidth]{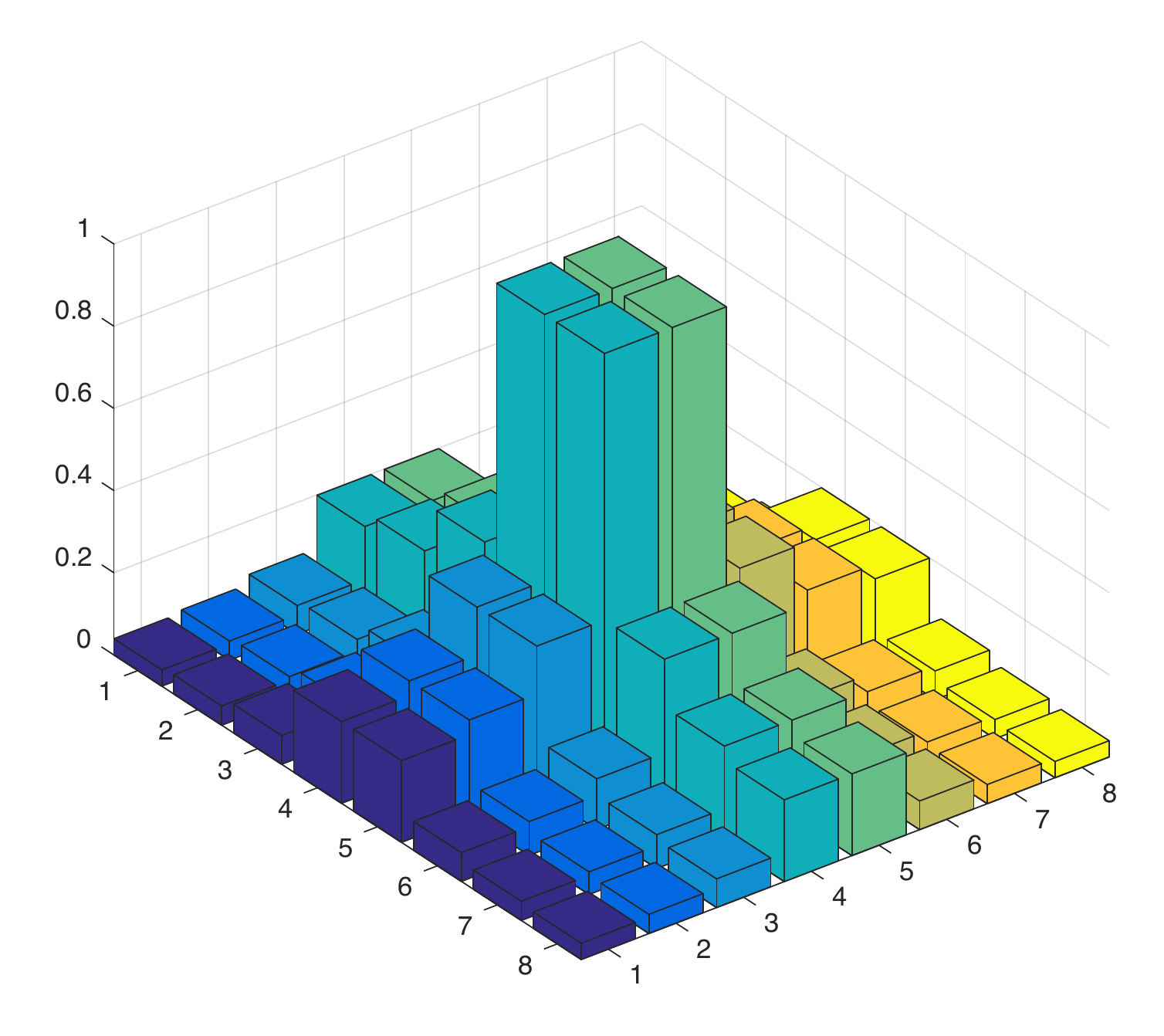}
        \caption{Normalized spatial spectrum for $8$x$8$ MIMO channel}
        \label{fig:fourtransforma}
    \end{subfigure}%
    ~
    \begin{subfigure}[t]{0.45\textwidth}
        \centering
        \includegraphics[width=\textwidth]{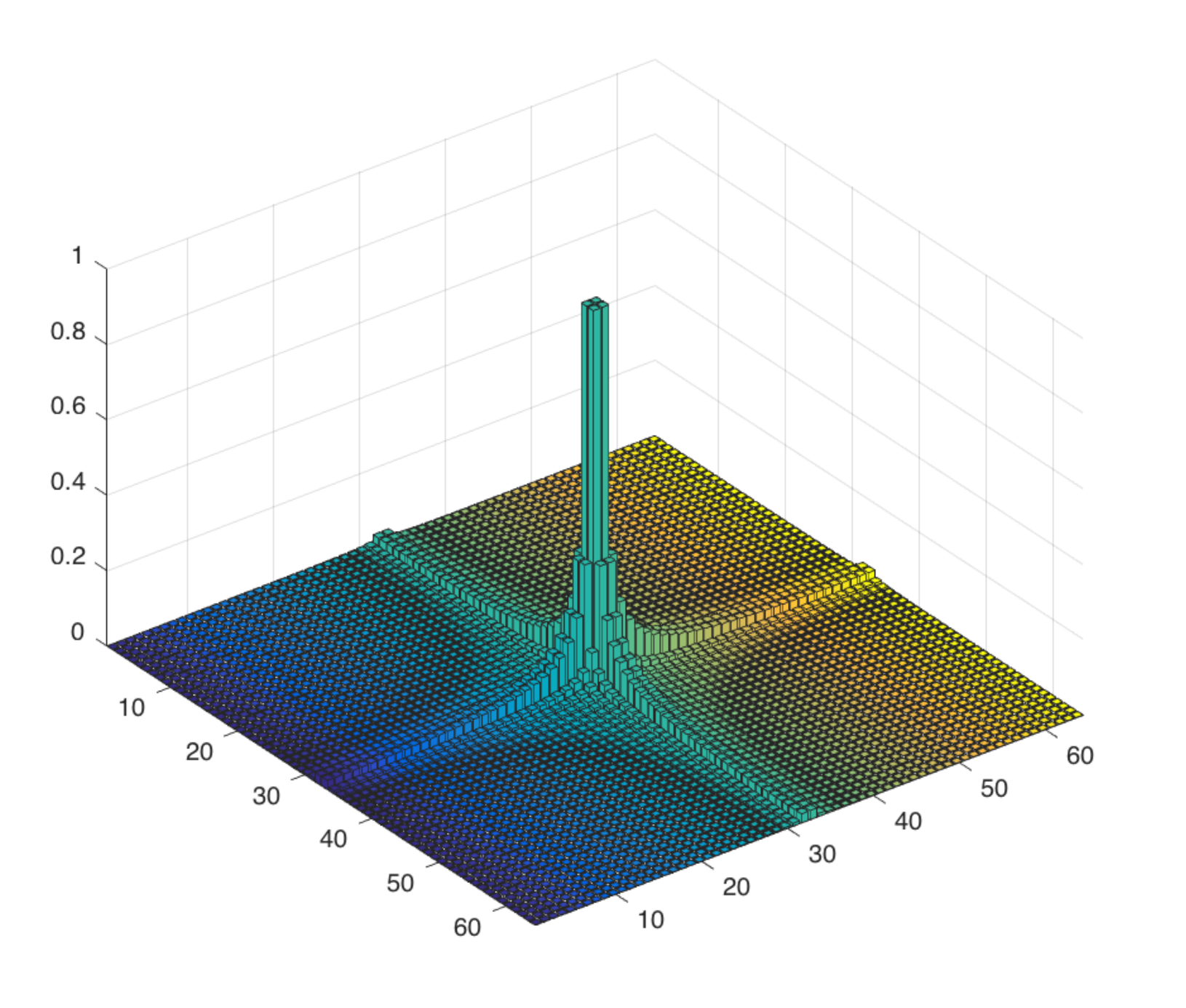}
        \caption{Normalized spatial spectrum for $64$x$64$ MIMO channel}
        \label{fig:fourtransformb}
    \end{subfigure}
\caption{Normalized spatial spectrum for $8$x$8$ and $64$x$64$ MIMO channels.}
\label{fig:fourtransform}
\end{figure}

In the \subsGHz~channel training phase, the TX transmits $\Ntrg\geq\MTXg$ training vectors. By collecting the training vectors in a matrix $\ul\bT=[\btg_1~\btg_2~\cdots~\btg_{\Ntrg}]$, we can write the collective received \subsGHz~signal as $\bRg=\bHg~\!\ul\bT+\ul\bV$, where $\ul\bV$ is the $\MRXg\times \Ntrg$ noise matrix with IID entries $[\ul\bV]_{m,n}\sim\cC\cN(0,\noisepg)$ . The least squares estimate of the channel is $\hat{\bHg}=\bRg~\!{\ul\bT}^{\dagger}$, where ${\ul\bT}^{\dagger}=\ul\bT^\ast (\ul\bT~\!\ul\bT^\ast)^{-1}$ is the pseudo inverse of $\ul\bT$. If the \subsGHz~channel correlation matrix and the noise variance are known at the receiver, the least squares estimate can be replaced with the minimum mean squared error estimate. Using the channel estimate, we get
\begin{align}
\hat{\bGg}=\ARXg^\ast\hat{\bHg}~\!\ATXg,
\label{eq:beamspacesubsghz}
\end{align}
where $\ARXg$ is the $\MRXg\times\MRXg$ DFT matrix and $\ATXg$ is the $\MTXg\times\MTXg$ DFT matrix. We refer to $|\hat\bGg|$ as the spatial spectrum. Example normalized spatial spectrums for $8\times 8$ and $64\times 64$ MIMO channels are shown in Fig.~\ref{fig:fourtransform}. We suppose that the $8\times 8$ channel corresponds to \subsGHz, and the $64 \times 64$ channel corresponds to mmWave. The figure shows that the Fourier analysis on \subsGHz~channel can give a coarse estimate of the directional information at mmWave.

The spatial spectrum $|\bGg|$ (the $\hat{}$ accent is removed for notational convenience) is directly used in weighted sparse signal recovery. The structured random codebook design, however, requires the indices of the dominant \subsGHz~AoA and AoD directions. The indices of the dominant direction can be found as
\begin{align}
\{\ul{i}^\star,\ul{j}^\star\}=\underset{1 \leq \ul{i} \leq \MRXg,1\leq \ul{j} \leq \MTXg}{\arg\max} |[\bG]_{\ul{i},\ul{j}}|.
\label{eq:dominantindices}
\end{align}
\subsubsection{Second stage (OOB-aided compressed beam-selection)}\label{subsec:WPDN}
We explain the OOB-aided compressed beam-selection in three parts. The first part is the weighted sparse recovery using information from a single subcarrier, the second is the structured random codebook design, and the third is the joint weighted sparse recovery using information from all active subcarriers.

\textbf{Weighted sparse recovery:} The OMP based sparse recovery assumes that the prior probability of the support is uniform, i.e., all elements of the unknown can be active with the same probability $p$. If some prior information about the non-uniformity in the support is available, the OMP algorithm can be modified to incorporate this prior information. In~\cite{Scarlett2013Compressed} a modified OMP algorithm called logit weighted - OMP (LW-OMP) was proposed for non-uniform prior probabilities. Assume that $\bp \in \bbR^{\MTX\MRX}$ is the vector of prior probabilities. Specifically, the $r$th element of $\bsfg[k]$ can be active with prior probability $0 \leq [\bp]_r \leq 1$. Then $r^\star$ can be found using LW-OMP as
\begin{align}
r^\star=\underset{1\leq r \leq \MRX\MTX}{\arg\max} |[\bPsi]_{:,r}^\ast\bsfy[k]| + w([\bp]_r),
\label{eq:OMPprojweighted}
\end{align}
where $w([\bp]_r)$ is an additive weighting function. The authors refer the interested reader to~\cite{Scarlett2013Compressed} for the details of LW-OMP and the selection of $w([\bp]_r)$. The general form of $w([\bp]_r)$ can be given as $w([\bp]_r)=J_{\rmw} \log\dfrac{[\bp]_r}{1-[\bp]_r}$, where $J_{\rmw}$ is a constant that depends on sparsity level, the amplitude of the unknown coefficients, and the noise level. In the absence of prior information, \eqref{eq:OMPprojweighted} can be solved using uniform probability $\bp=\delta\bone$, where $0<\delta<=1$, which is equivalent to solving~\eqref{eq:OMPproj}.

\begin{figure}[h!]
\centering
\includegraphics[width=0.45\textwidth]{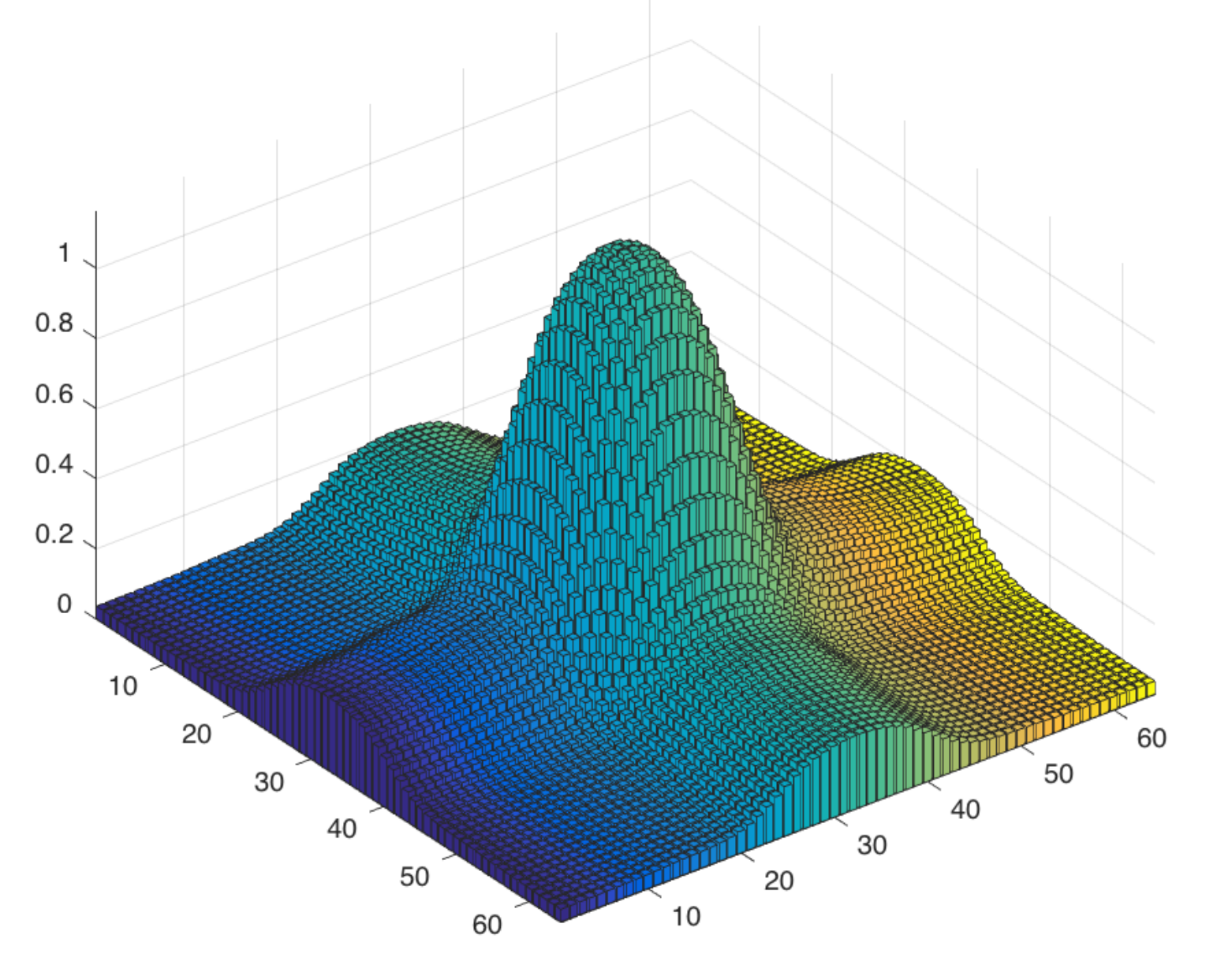}
\caption{The $64$x$64$ scaled spatial spectrum of the $8$x$8$ spatial spectrum in Fig.~\ref{fig:fourtransforma}.}
\label{fig:fourtransformc}
\end{figure}

The spatial information from \subsGHz~can be used to obtain a proxy for $\bp$. The probability vector $\bp\in\bbR^{\MRX\MTX}$ is obtained using $|\bGg|\in\bbR^{\MRXg\times\MTXg}$. To obtain $\bp$, we remove the dimensional discrepancy and find a scaled spatial spectrum $|\bGg_\rms|=g_{\rms}(|\bGg|)$, where $g_{\rms}(\cdot):\bbR^{\MRXg\times\MTXg}\rightarrow\bbR^{\MRX\times\MTX}$ is a two-dimensional scaling function. As an example, the scaling function can be implemented using bi-cubic interpolation~\cite{keys1981cubic}. The scaled spatial spectrum $|\bGg_\rms|$ corresponding to Fig.~\ref{fig:fourtransforma} is shown in Figure~\ref{fig:fourtransformc}.
A simple proxy of the probability vector based on the scaled spectrum $|\bGg_s|$ can be
\begin{align}
\bp=J_{\rmp} \dfrac{|\bgg_\rms-\min(\bgg_\rms)|}{\max(\bgg_\rms-\min(\bgg_\rms))},
\end{align}
where $J_\rmp$ is an appropriately chosen constant and $\bgg_\rms=\vec(\bGg_\rms)$. Initially the minimum of the spectrum $\bgg_\rms$ is subtracted to ensure all entries are non-negative. Then the spectrum is normalized to meet the probability constraint $0\leq[\bp]_r\leq 1$. The scaling constant $J_\bp\in(0,1]$ captures the reliability of the OOB-information. The reliability is a function of the \subsGHz~and  mmWave spatial congruence, and operating \SNR. For highly reliable information, a higher value can be used for $J_\rmp$.

\textbf{Structured random codebooks:} So far we have considered random codebooks that respect the analog hardware constraints, i.e., constant modulus and quantized phase-shifts. The random codebooks used for training, however, can be tailored to OOB information. We describe the design of structured codebooks for precoders, but it also applies to the combiners. 

Note that the beamspace representation of the channel divides the spatial AoD range into $\MTXg$ intervals of width $\frac{1}{\MTXg}$ each. We call these intervals angle bins. Recall from~\eqref{eq:dominantindices} that $\ul{j}^\star$ is the index associated with the dominant AoD. Thus, the interval $[\omegag_{\ul{j}^\star}-\frac{1}{2\MTXg},\omegag_{\ul{j}^\star}+\frac{1}{2\MTXg})$ is the estimated dominant \subsGHz~spatial angle bin. Due to a large number of antennas at mmWave in comparison with \subsGHz, the mmWave anlge bins have a smaller width, and several mmWave angle bins fall in one \subsGHz~angle bin. We create a set $\cJ$ using the mmWave angle bins that fall in the interval $[\omegag_{\ul{j}^\star}-\frac{1}{2\MTXg},\omegag_{\ul{j}^\star}+\frac{1}{2\MTXg})$, i.e., $\cJ=\{\phi_j: \phi_j=\arcsin(\frac{\omega_j}{d}), \omega_j\in[\omegag_{\ul{j}^\star}-\frac{1}{2\MTXg},\omegag_{\ul{j}^\star}+\frac{1}{2\MTXg}), j\in [\MTX]\}$. Using the set $\cJ$, we obtain a deterministic codebook $\bD=\aTX(\cJ)$, where $\aTX(\cJ)$ is the array response vector $\aTX$ evaluated at the elements of $\cJ$. Next, we construct a super-codebook $\bar\bF$ containing $\NTXb$ codewords according to~\cite{Alkhateeb2015Compressed}. The desired codebook then consists of the $\NTX$ codewords from the super codebook that have the highest correlation with the deterministic codebook $\bD$. The procedure to generate structured precoding codebooks is summarized in Algorithm~\ref{alg:struc_codebook}. The LW-OMP algorithm with structured codebooks is referred to as structured LW-OMP.

\begin{algorithm}
\caption{Structured random codebook design}
\begin{algorithmic}[1]
\INPUT $\ul{j}^\star$
\OUTPUT $\bF$
\STATE Construct a deterministic codebook $\bD=\aTX(\cJ)$, where  $\cJ=\{\phi_j: \phi_j=\arcsin(\frac{\omega_j}{d}), \omega_j\in[\omegag_{\ul{j}^\star}-\frac{1}{2\MTXg},\omegag_{\ul{j}^\star}+\frac{1}{2\MTXg}), j\in [\MTX]\}$.
\STATE Construct a super-codebook $\bar\bF$ using $\NTXb$ random codewords generated according to~\cite{Alkhateeb2015Compressed}.
\STATE Let $\bN=\bar\bF^\ast\bD$. Populate the index set $\cM$ with the indices of $\NTX$ rows of $\bN$ that have the largest 2-norms.
\STATE Create the precoding matrix $\bF=[\bar\bF]_{:,\cM}$.
\end{algorithmic}
\label{alg:struc_codebook}
\end{algorithm}
\textbf{Joint weighted sparse recovery:} If the unknowns $\bsfg[k]$ were recovered on all subcarriers, a suitable criterion for choosing $r^\star$ could be $r^\star=\underset{1\leq r \leq \MRX\MTX}{\arg\max} \sum_{k\in[K]}|[\bsfg[k]]_r|$. One can recover the vectors $\bsfg[k]~\forall k\in K$ individually on each subcarrier and then find $r^\star$. Instead, we note that the unknown sparse vectors have a similar sparsity pattern i.e., they share an approximately common support~\cite{Gao2015Spatially}. To exploit the common support property, we formulate the problem of recovering $\bsfg[k]~\forall k\in K$ jointly using measurements from all subcarriers. Formally, we collect all vectors $\bsfy[k]~\forall k\in[K]$ in a matrix $\bar\bsfY$, which can be written as
\begin{align}
\bar\bsfY&=\left[\bsfy[1]~\bsfy[2]~\cdots~\bsfy[K]\right],\nonumber\\
&=\bPsi \left[\bsfg[1]~\bsfg[2]~\cdots~\bsfg[K]\right]+[\bsfv[1]~\bsfv[2]~\cdots~\bsfv[K]]\nonumber,\\
&=\bPsi\bar\bsfG+\bar\bsfV.
\label{eq:MMVequation}
\end{align}

The columns of $\bar\bsfG$ are approximately jointly sparse, i.e., $\bar\bsfG$ has only a few non-zero rows. The sparse recovery problems of the form \eqref{eq:MMVequation} are referred to as MMV problems. The simultaneous OMP (SOMP) algorithm~\cite{Tropp2005Simultaneous} is an OMP variant tailored for MMV problems. Using SOMP, $r^\star$ can be found as
\begin{align}
r^\star=\underset{1\leq r \leq \MRX\MTX}{\arg\max}  \sum_{k\in [K]}  |[\bPsi]_{:,r}^\ast\bsfy[k]|.
\label{eq:SOMPproj}
\end{align}

The OOB information can be incorporated in SOMP algorithm via logit weighting and structured codebooks. Specifically, the logit weighted - SOMP (LW-SOMP) algorithm~\cite{Li2014Compressed} finds $r^\star$ by
\begin{align}
r^\star=\underset{1\leq r \leq \MRX\MTX}{\arg\max}  \sum_{k\in [K]}  |[\bPsi]_{:,r}^\ast\bsfy[k]| +w([\bp]_r).
\label{eq:SOMPprojweighted}
\end{align}
The LW-SOMP algorithm used with structured random codebooks is termed structured LW-SOMP.
\section{Simulation Results}\label{sec:simres}
In this section, we present simulation results to test the performance of the proposed OOB-aided mmWave beam-selection strategies. The \subsGHz~system operates at $\fg=3.5~\GHz$ with $1~\MHz$ bandwidth and $\MRXg=\MTXg=4$ antennas. The mmWave system operates at $f=28~\GHz$ with $320~\MHz$ bandwidth and $\MRX=\MTX=32$ antennas. Both systems use ULAs with half wavelength spacing $\dg=d=1/2$. The number of OFDM subcarriers is $K=256$, and the CP length is  $\Lc=K/4=64$. With the chosen operating frequencies, the number of antennas, and the inter-element spacing, the array aperture for \subsGHz~and mmWave arrays is the same. Further, as $K+\Lc=320$, the chosen \subsGHz~and mmWave bandwidths imply that the mmWave OFDM block has the same time duration as the \subsGHz~symbol. The transmission power for \subsGHz~and mmWave systems is $\Ptg=\Pt=37~\dBm$, and the path-loss coefficient at \subsGHz~and mmWave is $3$. The \subsGHz~channel is frequency flat, whereas the mmWave channel is frequency selective with $L=63$ taps. The number of quantization bits for analog phase-shifters are $\DRX=\DTX=\log_2(32)=5$ to realize the DFT codebooks. The raised cosine filter with a roll off factor of $1$ is used as a pulse shaping filter.

To study the performance of OOB-aided compressed beam-selection, we fix the channel parameters of \subsGHz~and mmWave in the first experiment. The performance of OOB-aided beam-selection with varying degrees of spatial congruence between \subsGHz~and mmWave is studied in the second experiment. The \subsGHz~and mmWave channels have $\Cg=4$ and $C=3$ clusters respectively, each contributing $\Rcg=\Rc=10$ rays. We assume that the clusters are distributed uniformly in time and space. Hence, $\taucg \sim U[0,\taumaxg]$ and $\tauc \sim U[0,\taumax]$. As the maximum delay spread of  \subsGHz~channel is expected to be larger than the maximum delay spread of mmWave~\cite{weiler2015simultaneous,Poon2003Indoor,Jaeckel20165G,Kaya201628}, we choose $\taumaxg\approx 57~\ns$ and $\taumax\approx 48~\ns$ i.e., approximately $20\%$ larger maximum delay spread for \subsGHz. The mean AoAs/AoDs of the clusters are distributed as  $\{\ul{\theta}_c,\phig_c\}\sim U[-\frac{\pi}{2},\frac{\pi}{2})$ and $\{\thetac,\phic\}\sim U[-\frac{\pi}{2},\frac{\pi}{2})$. The relative time delays of the paths within the clusters are drawn from zero mean uniform distributions with RMS AS $\ul{\sigma}_{\ul{\tau}_{r_{\ul{c}}}}=\frac{\ul{\tau}_{\max}}{20\sqrt{12}}$ and $\sigma_{\tau_{r_c}}=\frac{\tau_{\max}}{20\sqrt{12}}$. The relative AoA/AoD shifts come from zero mean uniform distributions with AS $\{\ul{\sigma}_{\varthetag_{\cg}},\ul{\sigma}_{\varphig_{\cg}}\}=2.4^\circ\approx\SI{0.042}{\radian}$ and $\{\sigma_{\vartheta_c},\sigma_{\varphi_c}\}=2^\circ\approx\SI{0.035}{\radian}$. We use the replacement and perturbation models described in Section~\ref{sec:multi-bandch} with the angle modifier adjusted to limit the angles in $[-\frac{\pi}{2},\frac{\pi}{2})$.

The metric used for performance comparison is the effective achievable rate $\Reff$ defined as
\begin{align}
\Reff=\dfrac{\eta}{EK}\sum_{e=1}^{E}\sum_{k=1}^{K}\log_2\left(1+\frac{\Pt}{K\noisepsf}|\aRX^\ast(\nu_{\hat i})\bsfH[k]\aTX(\omega_{\hat j})|^2\right),
\end{align}
where $\{\hat{i}, \hat{j}\}$ are the estimated transmit and receive codeword indices, $E$ is the number of independent trials for ensemble averaging, $\eta\triangleq \max(0,1-\frac{\NRX\times\NTX}{\Tc})$, and $\Tc$ is the channel coherence time in OFDM blocks. With the channel coherence of $\Tc$ blocks and a training of $\NRX\times\NTX$ blocks, $1-\frac{\NRX\times\NTX}{\Tc}$ is the fraction of time/blocks that are used for data transmission. Thus, $\eta$ captures the loss in achievable rate due to the training.
\subsection{OOB-aided compressed beam-selection}\label{subsec:WCSandSD}
\begin{figure}[h!]
\centering
\includegraphics[width=0.45\textwidth]{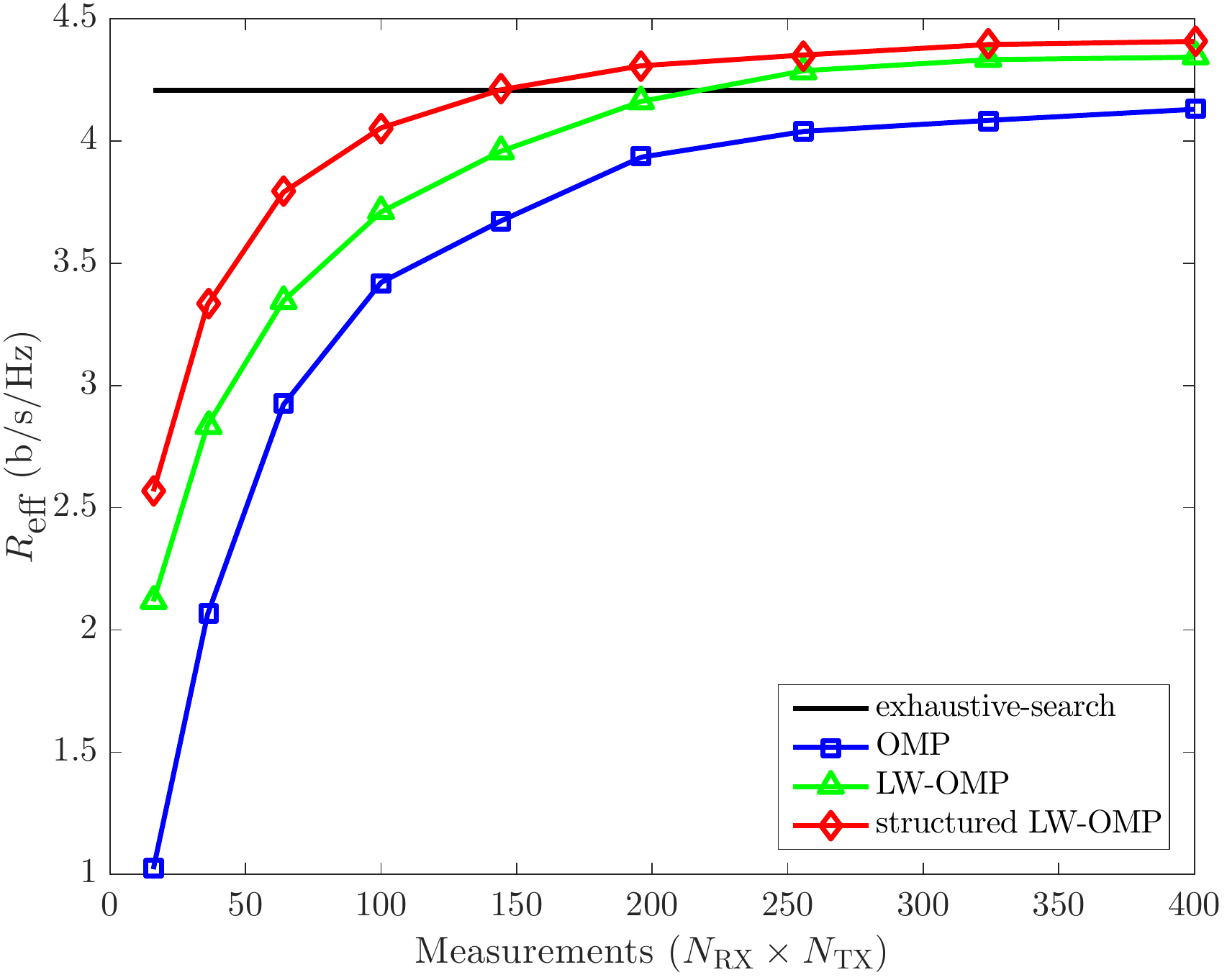}
\caption{Effective rate of the structured LW-OMP approach versus the number of measurements $\NRX\times\NTX$ with $\SI{40}{\meter}$ TX-RX separation and $\Tc=6(\MRX\times\MTX)=6144$ blocks.}
\label{fig:WCSWDCS}
\end{figure}

In this experiment, we test the performance of OOB-aided compressed beam-selection in comparison with in-band only compressed beam-selection. The TX-RX separation for this experiment is fixed at $\SI{40}{\meter}$. The compressed beam-selection is performed using information on a single subcarrier, chosen uniformly at random from the $K$ subcarriers. The number of independent trials is $E=2000$. The number of measurements for exhaustive-search are fixed at $32\times32=1024$. The rate results as a function of the number of measurements $\NRX\times\NTX$ are shown in Fig.~\ref{fig:WCSWDCS}. It can be observed that throughout the range of interest the OOB-aided compressed beam-selection using LW-OMP has a better effective rate in comparison with OMP. Further structured LW-OMP improves on the performance of LW-OMP, validating the benefit of using structured random codebooks.

\begin{figure}[h!]
\centering
\includegraphics[width=0.45\textwidth]{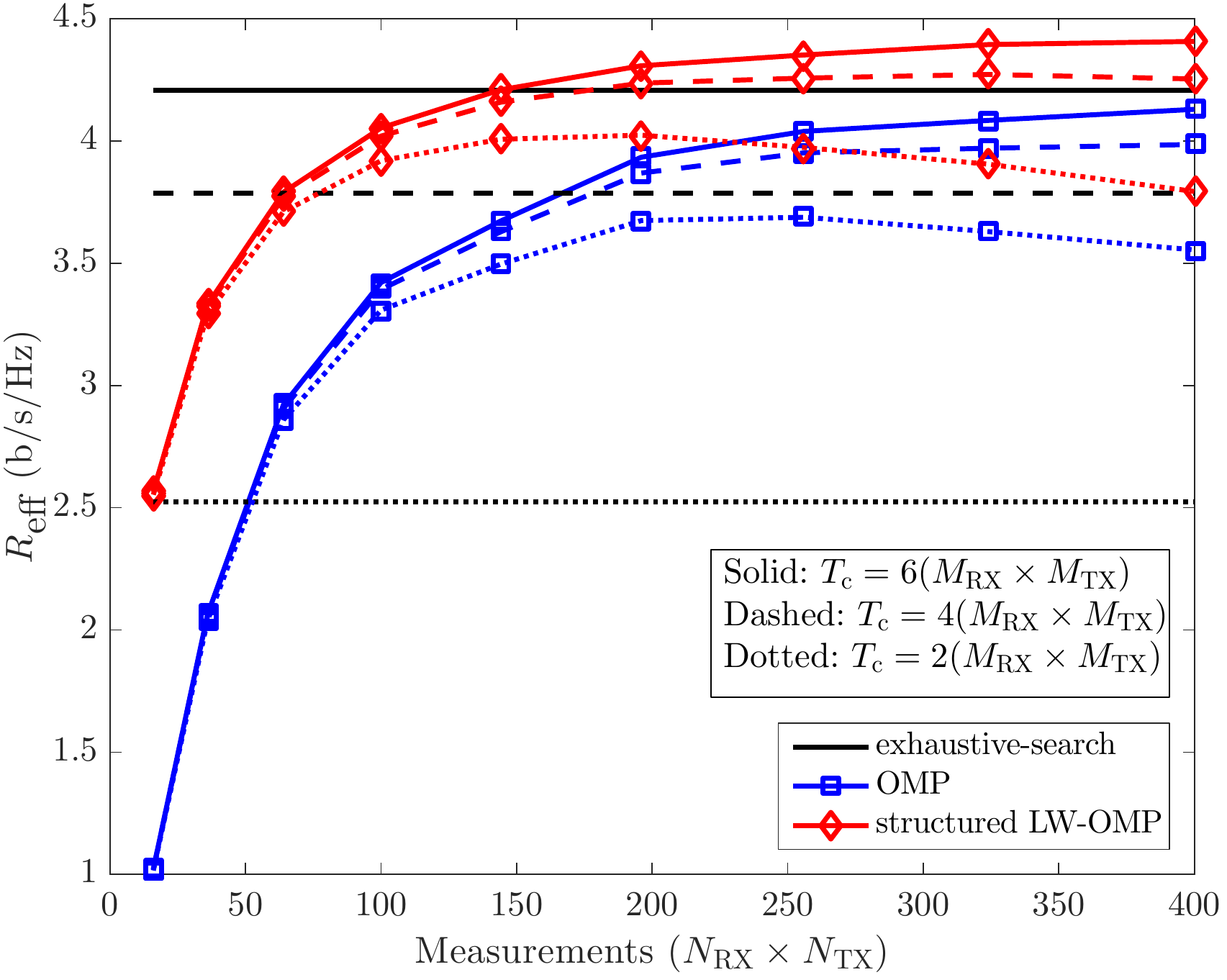}
\caption{Effective rate of the structured LW-OMP approach versus the number of measurements $\NRX\times\NTX$ with $\SI{40}{\meter}$ TX-RX separation and three different channel coherence times $\Tc$.}
\label{fig:Alldiffcoh}
\end{figure}

It is observed from Fig.~\ref{fig:WCSWDCS} that the effective rate of structured LW-OMP only slightly improves on the rate of exhaustive-search. This, however, is true for large channel coherence $\Tc$ values. We plot the effective rate of the proposed structured LW-OMP based compressed beam-selection for three channel coherence values in Fig.~\ref{fig:Alldiffcoh}. As the coherence time of the channel decreases, the advantage of the proposed approach becomes significant. As an example, for a medium channel coherence time i.e., $4(\MRX\times\MTX)$, the proposed structured LW-OMP based compressed beam-selection can reduce the training overhead of exhaustive-search by over $20$x and the overhead of LW-OMP by $4$x. The gains for smaller channel coherence times are more pronounced. Therefore, the proposed approach is suitable for applications with rapidly varying channels e.g., V2X communications.

\begin{figure}[h!]
\centering
\includegraphics[width=0.45\textwidth]{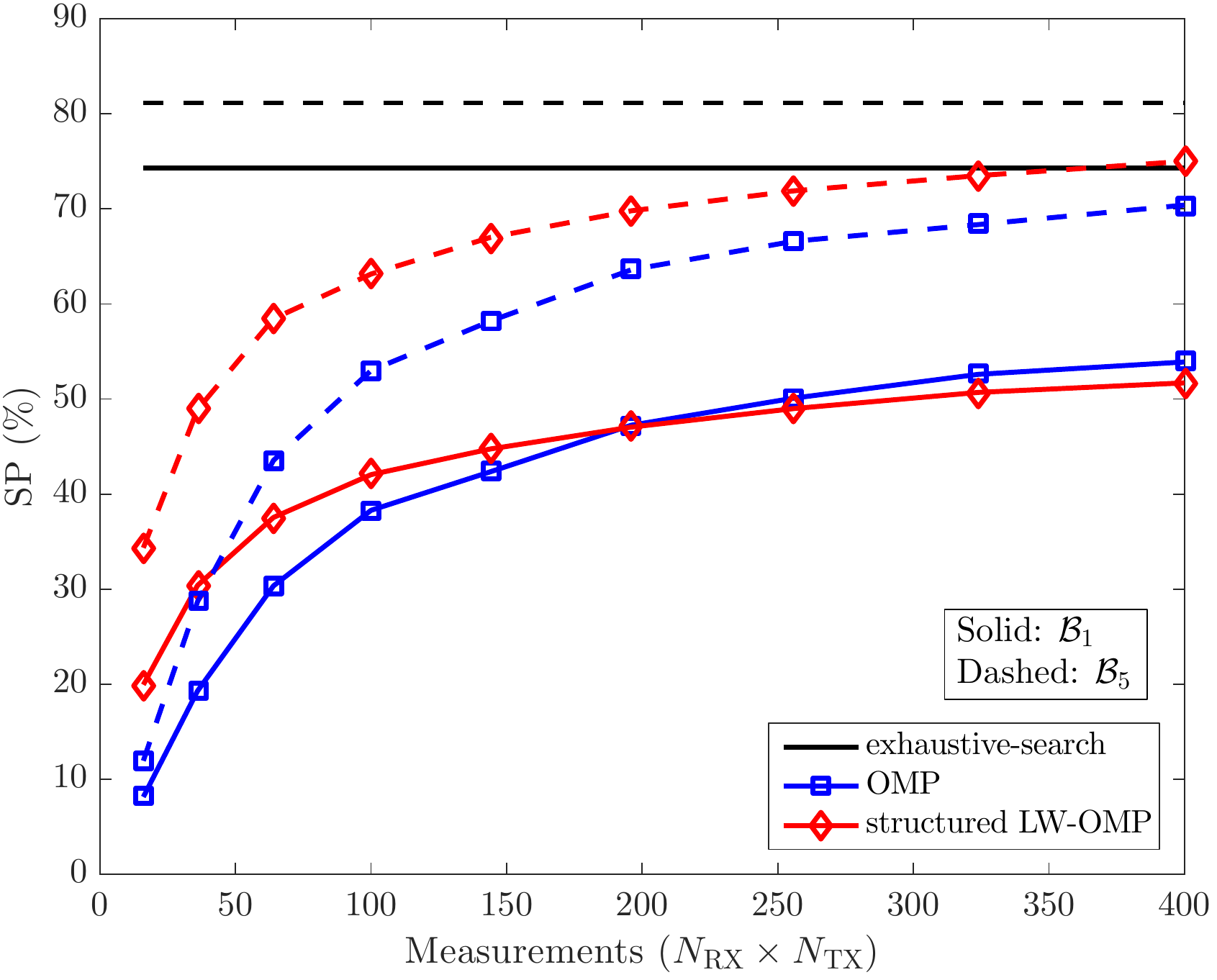}
\caption{Success percentage of the structured LW-OMP approach versus the number of measurements $\NRX\times\NTX$ with $\SI{40}{\meter}$ TX-RX separation.}
\label{fig:Succ_Perc}
\end{figure}

To study the fraction of times the proposed approach recovers the best beam-pair, we define and evaluate the success percentage of the proposed approach. The success percentage is defined as
\begin{align}
\mathrm{SP}=\dfrac{1}{E}\sum_{e=1}^E |\hat r^\star \cap \cB_N|,
\end{align}
where $\hat r^\star$ is the index estimated by the proposed approach and $\cB_N$ is the set containing the actual indices corresponding to the $N$ best TX/RX beam-pairs. When $N=1$, the set $\cB_1$ has only one element and that is the index corresponding to the beam-pair with the highest receive power. For $N>1$, the set has $N$ entries that are indices corresponding to the $N$ beam-pairs with the highest receive power. Using a set of indices, instead of the index corresponding to the best beam-pair, generalizes the study and reveals an interesting behavior about selecting one of the better beam-pairs in comparison with selecting the best beam-pair. For now, note that due to grid offset and the presence of multiple clusters in the mmWave channel, it is possible that the proposed approach does not recover the best beam-pair and still manages to provide a decent effective rate. We populate the set $\cB_N$ by performing exhaustive-search in a noiseless channel. We do so as the exhaustive-search in a noisy channel is itself subject to errors. This behavior is revealed in Fig.~\ref{fig:Succ_Perc}, where the exhaustive-search succeeds $\approx 74\%$ of the times for $\cB_1$ and $\approx 80 \%$ of the times for $\cB_5$. The success percentage of the proposed structured LW-OMP algorithm is $\approx 50\%$ for $\cB_1$ and $\approx 75\%$ for $\cB_5$. The high success percentage for $\cB_5$ is a ramification of having several strong candidate beam-pairs due to grid offset and the presence of multiple clusters. Note that even though the proposed approach has a (slightly) inferior success percentage for $\cB_5$ compared with the exhaustive-search, the training overhead of the proposed approach is significantly lower. With the overhead factored in, the proposed approach is advantageous compared to exhaustive-search as evidenced by the effective rate results in Fig.~\ref{fig:Alldiffcoh}.

\begin{figure}[h!]
\centering
\includegraphics[width=0.45\textwidth]{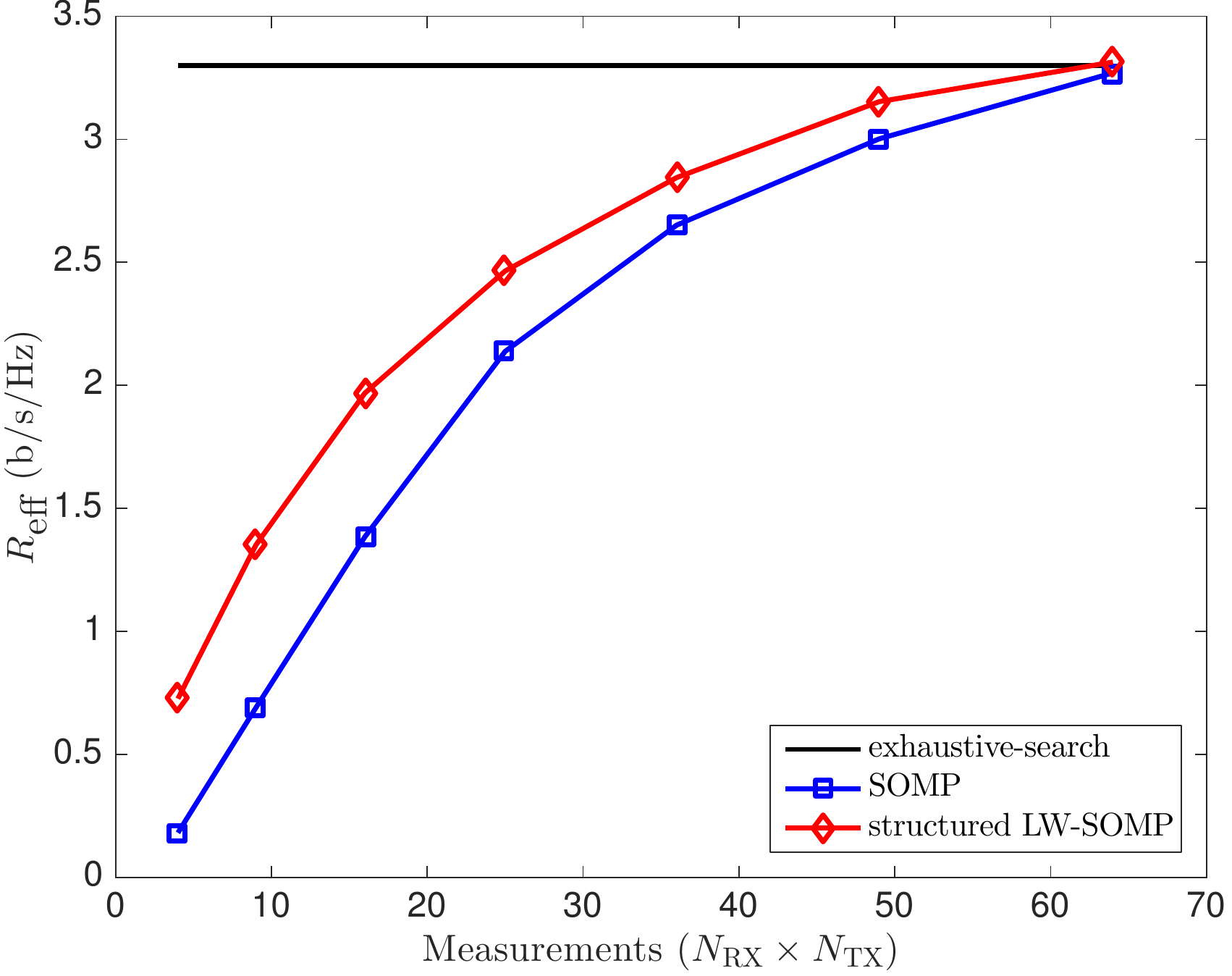}
\caption{Effective rate of the structured LW-SOMP approach versus the number of measurements $\NRX\times\NTX$ with $\SI{80}{\meter}$ TX-RX separation and $\Tc=6(\MRX\times\MTX)=6144$ blocks.}
\label{fig:MMV}
\end{figure}

Next, we evaluate the performance of structured LW-SOMP based compressed beam-selection using information from all active subcarriers. The TX-RX separation is $\SI{80}{\meter}$, and with this separation the pre-beamforming per subcarrier \SNR~is $\approx\SI{-10}{\decibel}$. The results of this experiment are shown in Fig.~\ref{fig:MMV}. The structured LW-SOMP achieves a better effective rate in comparison with LW-SOMP. Due to the use of training information from all subcarriers, both structured LW-SOMP and LW-SOMP reach the effective rate of exhaustive-search with a handful of measurements. For low channel coherence times $\Tc$, the compressed beam-selection approaches, especially OOB-aided compressed beam-selection, will outperform exhaustive-search.

\subsection{Impact of mismatch in spatial parameters}\label{sec:ch_mis}
The performance of OOB-aided approaches is a function of the congruence between \subsGHz~and mmWave channels. We choose structured LW-OMP as a representative OOB-aided beam-selection approach and test the effective rate of structured LW-OMP with varying degrees of congruence between \subsGHz~and mmWave. It is worth highlighting that in the proposed OOB-aided beam-selection strategies only spatial information of the \subsGHz~channels is used. As such mismatch in the time parameters of \subsGHz~and mmWave is not expected to impact the performance of OOB-aided approaches. Hence, we test the performance of structured LW-OMP as a function of mismatch in the spatial parameters. Specifically, we assume a single cluster in \subsGHz~and mmWave i.e., $\Cg=1$ and $C=1$ and study the impact of mismatch in the AoA/AoD and AS of the \subsGHz~and mmWave clusters. The channel parameters not explicitly mentioned in this experiment are consistent with the previously used setup. For this experiment, the number of measurements is $\NTX\times\NRX=64$, the coherence time is $\Tc=\infty$, and the number of independent trials is $E=5000$.

\begin{figure}[h!]
\centering
\includegraphics[width=0.45\textwidth]{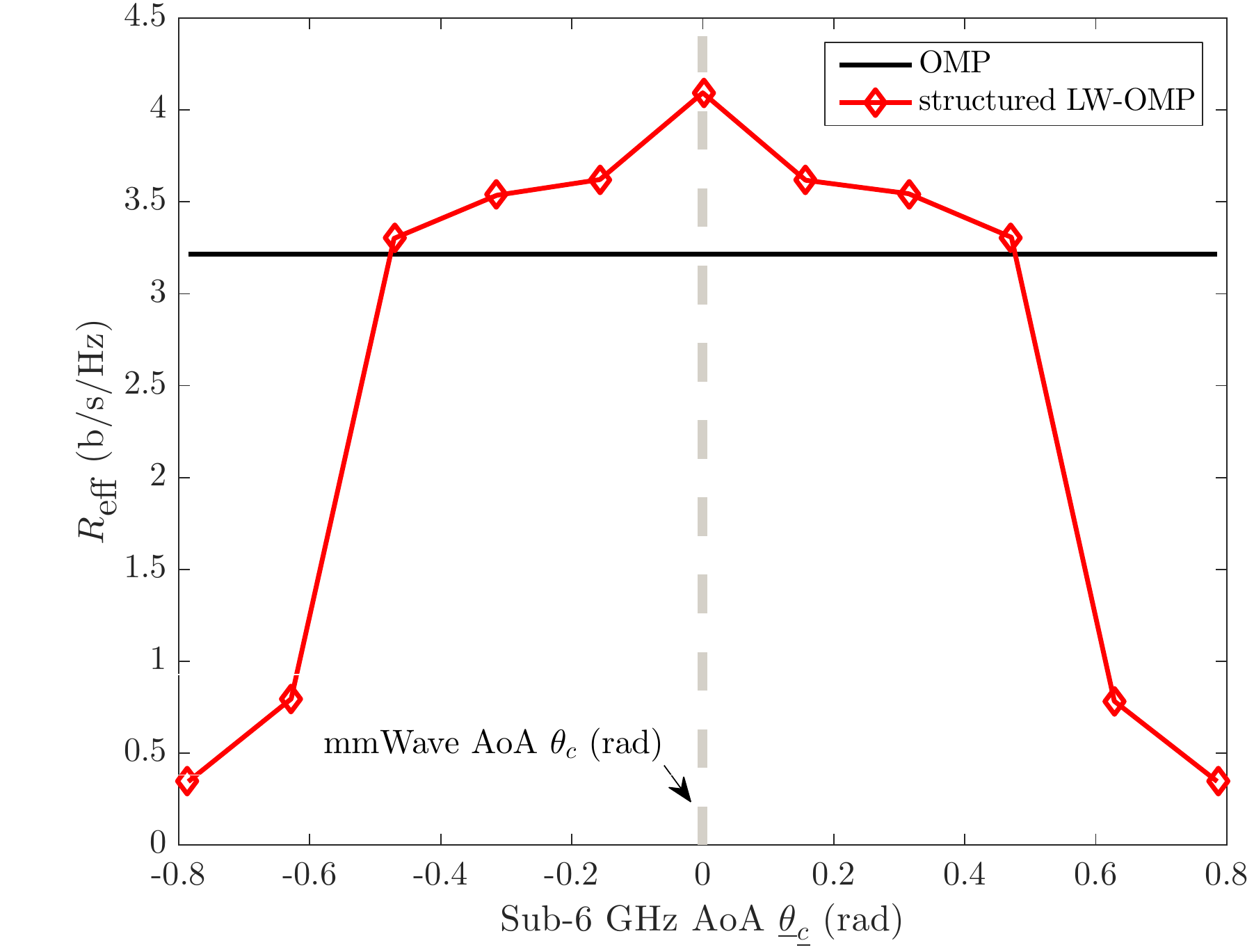}
\caption{Effective rate of the structured LW-OMP approach versus \subsGHz~AoA with $\SI{200}{\meter}$ TX-RX separation.}
\label{fig:Angle_Sep}
\end{figure}

\begin{figure}[h!]
\centering
\includegraphics[width=0.45\textwidth]{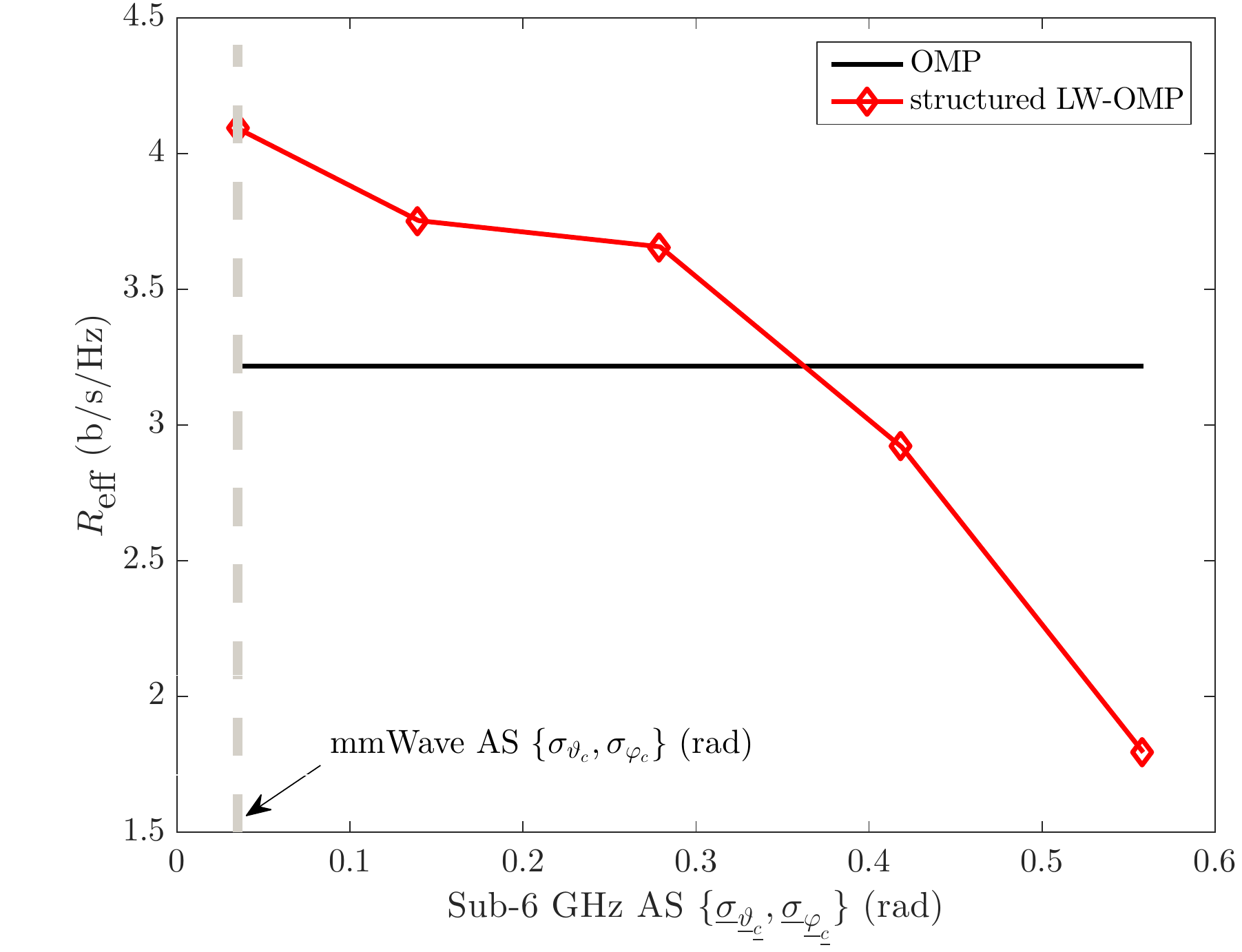}
\caption{Effective rate of the structured LW-OMP approach versus \subsGHz~AS with $\SI{200}{\meter}$ TX-RX separation.}
\label{fig:AS_var}
\end{figure}

We study the impact of mismatch between the mean AoA/AoD of the \subsGHz~and mmWave cluster in Fig.~\ref{fig:Angle_Sep}. We keep the mean AoA/AoD of the mmWave cluster fixed at $\{\thetac,\phic\}=\{0,0\}\SI{}{\radian}$. We also keep the mean AoD of \subsGHz~fixed at $\phig_{\cg}=\SI{0}{\radian}$ and vary the mean AoA $\thetagg_{\cg}$. Expectedly when the $\thetagg_{\cg}$ matches $\thetac$, we see the highest rate for structured LW-OMP. As $\thetagg_{\cg}$ differs from $\thetac$ by more than $\approx \SI{0.5}{\radian}$, the structured LW-OMP becomes inferior to in-band only OMP based beam-selection. Keeping the AoA $\thetagg_{\cg}$ fixed and varying the AoD $\phig_{\cg}$ is expected to yield similar results.

We also study the impact of mismatch between the AS of the \subsGHz~and mmWave cluster. We keep the AoA/AoD of \subsGHz~and mmWave fixed at $\{\thetagg_{\cg},\phig_{\cg}\}=\{0,0\}\SI{}{\radian}$ and $\{\thetac,\phic\}=\{0,0\}\SI{}{\radian}$. The relative AoA/AoD shifts of the paths within mmWave cluster have zero mean, uniform distribution and AS $\{\sigma_{\vartheta_c},\sigma_{\varphi_c}\}=2^\circ\approx\SI{0.035}{\radian}$. The relative AoA/AoD shifts of the paths within \subsGHz~cluster are also zero man and uniformly distributed. We test the performance of the structured LW-OMP algorithm for varying AS of the \subsGHz~cluster. Specifically, we increase the AS of \subsGHz~cluster compared to mmWave, which is in accordance with the observations of prior work~\cite{Kaya201628,Poon2003Indoor}. Both the AoA and AoD AS are increased equally. The effective rate is plotted in Fig.~\ref{fig:AS_var}. As expected, the gain of structured LW-OMP decreases as the AS of~\subsGHz~increases. In this particular experiment, when the AS of \subsGHz~was $\approx\SI{0.368}{\radian}$, i.e., $\approx 10.5$x the mmWave AS, the performance of structured LW-OMP became inferior to the in-band only OMP.
\section{Conclusion}\label{sec:conc}
In this paper, we used the \subsGHz~spatial information to reduce the training overhead of beam-selection in an analog mmWave system. We formulated the compressed beam-selection problem as a weighted sparse recovery problem with structured random codebooks to incorporate out-of-band information. We evaluated the achievable rate of the proposed out-of-band aided beam-selection strategies using multi-band frequency dependent channels. The channels were generated using the proposed multi-band channel generation methodology that is consistent with the frequency dependent channel behavior observed in the prior work. From the rate results, it was concluded that the training overhead of in-band only compressed beam-selection can be reduced by $4$x if the out-of-band information is used. 

There are several directions for future work. The frequency dependent channel estimation strategy can be calibrated with the emerging joint channel modeling results for \subsGHz~and mmWave. The out-of-band aided beam-selection strategies can be explored for arrays other than uniform linear arrays, e.g., circular and planar arrays. Finally, using out-of-band information in hybrid analog/digital and fully digital low-resolution architectures for mmWave systems is an interesting direction for future work.

\bibliographystyle{IEEEtran}
\bibliography{\centrallocation/Abbr,\centrallocation/Master_Bibliography}{}
\end{document}